\documentclass[%
reprint,
groupedaddress,
 amsmath,amssymb,
 aps,
prb,
]{revtex4-2}

\usepackage {graphicx,epsfig,graphics,color}
\usepackage{dcolumn}
\usepackage{bm}
\usepackage{hyperref}
\usepackage{sidecap}
\usepackage{float}

\begin{document}

\title{Magnetic field control over the axialness of Higgs modes in charge-density wave compounds}

\author{Dirk Wulferding${^{1,2*}}$, Jongho Park${^{1,2}}$, Takami Tohyama${^{3}}$, Seung Ryong Park${^{4}}$, and Changyoung Kim${^{1,2*}}$}

\address{${^1}$Center for Correlated Electron Systems, \mbox{Institute for Basic Science, Seoul 08826, Korea}\\
${^2}$Department of Physics and Astronomy, \mbox{Seoul National University, Seoul 08826, Korea}\\
${^3}$Department of Applied Physics, \mbox{Tokyo University of Science, Tokyo 125-8585, Japan}\\
${^4}$Department of Physics, \mbox{Incheon National University, Incheon 22012, Korea}\\
${^*}$Corresponding authors: dirwulfe@snu.ac.kr; changyoung@snu.ac.kr\\
}

\date{\today}

\begin{abstract}

Understanding how symmetry-breaking processes generate order out of disorder is among the most fundamental problems of nature. The scalar Higgs mode - a massive (quasi-) particle - is a key ingredient in these processes and emerges with the spontaneous breaking of a continuous symmetry. Its related exotic and elusive axial counterpart, a Boson with vector character, can be stabilized through the simultaneous breaking of multiple continuous symmetries. Here, we employ a magnetic field to tune the recently discovered axial Higgs-type charge-density wave amplitude modes in rare-earth tritellurides. We demonstrate a proportionality between the axial Higgs component and the applied field, and a 90$^{\circ}$ phase shift upon changing the direction of the B-field. This indicates that the axial character is directly related to magnetic degrees of freedom. Our approach opens up an in-situ control over the axialness of emergent Higgs modes.

\end{abstract}

\maketitle


\section{\label{intro}Introduction}

Charge-density wave (CDW) phases play an essential role in condensed matter physics, where they are closely linked to exotic phases and emergent phenomena, such as unconventional superconductivity~\cite{mielke-22}, topologically non-trivial electronic phases~\cite{gooth-19}, or generating novel electronic states along domain walls~\cite{sipos-08, joe-14}. On a more fundamental level, charge-density waves serve as valuable platforms to explore general concepts such as the (tunability of the) nature of phase transitions, quantum criticality, and the breaking of symmetries. Especially those CDWs that are linked to unconventional ordering processes may offer access to explore phenomena postulated for high-energy particle physics. A recent polarization-resolved Raman spectroscopic study of the uni-directional charge-density wave materials GdTe$_3$ and LaTe$_3$ revealed a remarkable two-fold ($A_2$) symmetry of the CDW amplitude mode at room temperature and zero magnetic fields. This low symmetry is rationalized by quantum pathway interference processes that uniquely occur in these rare-earth tritellurides with two distinct charge-density wave vectors. In contrast to other, conventional CDW materials, $R$Te$_3$ hosts two (nearly) degenerate nesting conditions, \textbf{q}$_{\mathrm{CDW}}$ and \textbf{c}$^*$-\textbf{q}$_{\mathrm{CDW}}$, connecting $p_x$--$p_x$ ($p_y$--$p_y$) bands of Te, and mixing $p_x$--$p_y$ ($p_y$--$p_x$) bands, respectively~\cite{brouet-08, eiter-13}. The resulting two-fold periodic CDW amplitudon with vector character has since been dubbed as an axial Higgs mode~\cite{wang-22}, i.e., a condensed matter analogon to a highly elusive elementary particle.

Two natural questions arise: what is the microscopic mechanism that generates axialness in $R$Te$_3$, and how can we control or tune the nature of the Higgs mode? Its axialness dictates the breaking of additional symmetries. We can conceive of several potentially relevant scenarios, illustrated in Fig. 1a: (i) Since GdTe$_3$ orders antiferromagnetically below the N\'{e}el temperature $T_{\mathrm{N}} = 11.5$ K, a coupling between the Gd spins and the CDW may be a crucial component to the axial Higgs mode. (ii) A slight lattice distortion of the Te square-net units within the CDW phase may be conducive to a ferro-rotational state with axial properties. (iii) A finite orbital angular momentum (OAM) can be generated from the mixing of Te orbitals via $p_x \pm \mathrm{i}p_y$, which breaks time reversal symmetry and can respond to an external magnetic field. Motivated by this open issue, we performed a polarization-resolved Raman spectroscopy study on the two sister compounds GdTe$_3$ (with low-temperature antiferromagnetic order) and LaTe$_3$ (without any long-range magnetic order) at various temperatures and with applied magnetic fields. In this work, we show that the low two-fold symmetry of the Higgs mode persistently observed in both materials and across a wide range of temperatures, together with its dramatic field-dependence allows us to rule out spin degrees of freedom as a relevant ingredient, and ultimately highlights the relevance of orbital degrees of freedom to stabilize axial Higgs modes in $R$Te$_3$.

\section{\label{results}Results}

\subsection{\label{field-tune}Field-tuning the Higgs mode}

\begin{figure*}
\includegraphics[width=13cm]{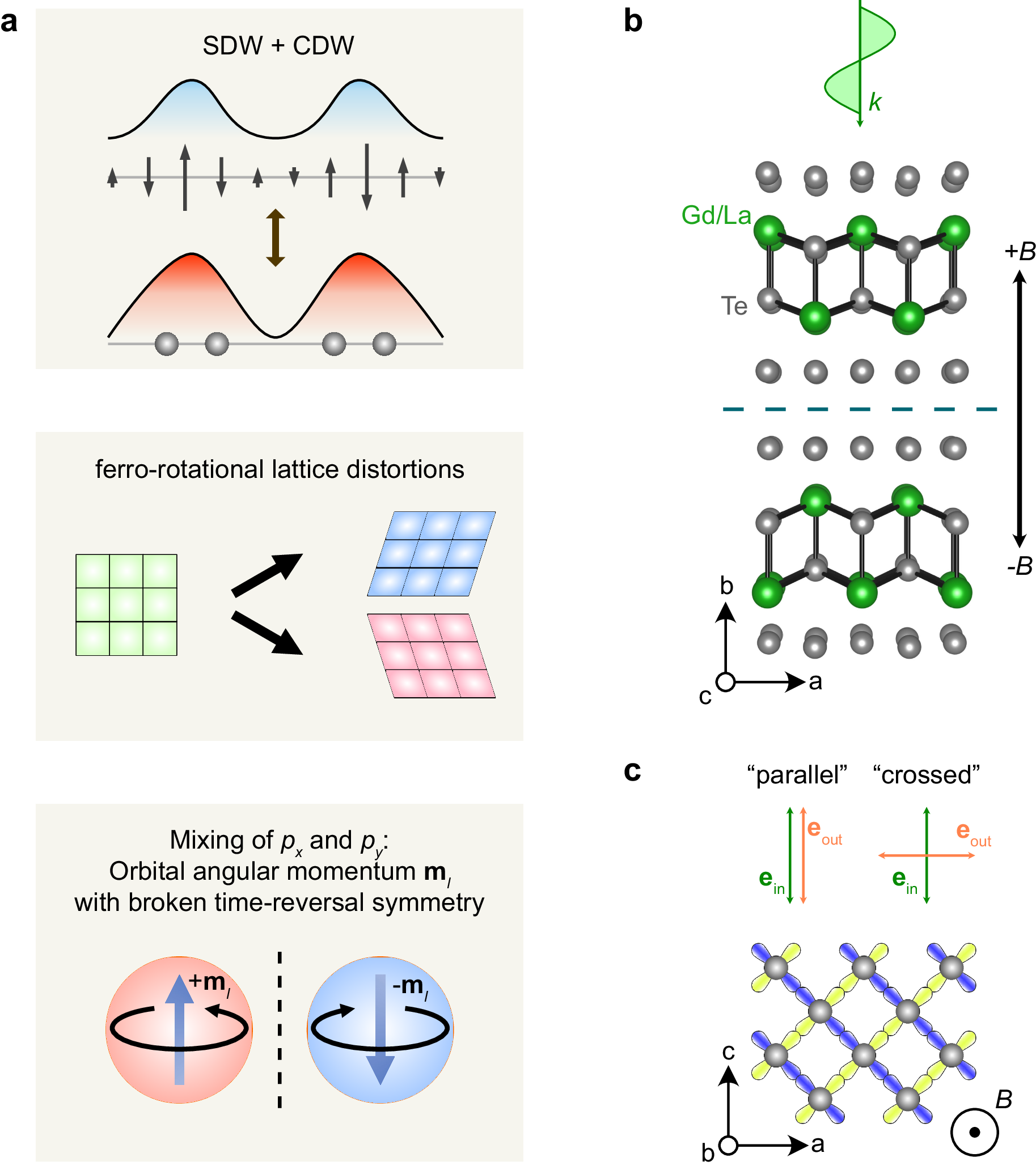}
\caption{\label{fig:1-schematics} \textbf{Symmetry breaking scenarios, crystal structure and scattering geometry.} \textbf{a}, Illustrations of potential symmetry breaking processes resulting in axial Higgs modes in GdTe$_3$: Coupling between magnetic order and charge-density wave below $T_{\mathrm{N}}$; ferro-rotational lattice distortion of Te square net units; mixing between $p_x$ and $p_y$ orbitals enabling a finite orbital angular momentum. \textbf{b}, Side view of alternating Te layers and slabs of GdTe (LaTe) stacked along the crystallographic $b$-axis. A natural cleaving plane exists between two van-der-Waals coupled Te layers (dashed line). An external magnetic field is applied out-of-plane, and the laser light (green arrow) propagates along the $b$-axis with its polarization within the $ac$-plane. \textbf{c}, Top-view ($ac$-plane) of the Te square lattice, with $p_x$ and $p_y$ orbitals drawn in blue and yellow. The configurations for parallel ($cc$) and crossed ($ca$) polarizations, later denoted as $\theta = 0^{\circ}$, are indicated by arrows.}
\end{figure*}

Figs. 1b and 1c outline the schematics of our experiment. Layers of [Te]--[GdTe]--[Te] building blocks are stacked along the crystallographic $b$-axis~\cite{norling-66}, with unidirectional CDW order emerging within the Te square lattice along the $a$ or the $c$ direction~\cite{dimasi-95}. The incident laser light direction and the magnetic field are both aligned out-of-plane, i.e., along the $b$-axis. We probe the excitations with light polarized within the $ac$-plane. The configurations sketched in Fig. 1c correspond to $cc$ (in parallel configuration) and to $ca$ (in crossed polarization), and are later on denoted as $\theta = 0^{\circ}$ (see below). Within this scheme we obtain highly polarization-resolved Raman data by continuously rotating the light polarization within the $ac$-plane, while keeping a fixed relation between \textbf{e}$_{\mathrm{in}}$ and \textbf{e}$_{\mathrm{out}}$.

\begin{figure*}
\includegraphics[width=14cm]{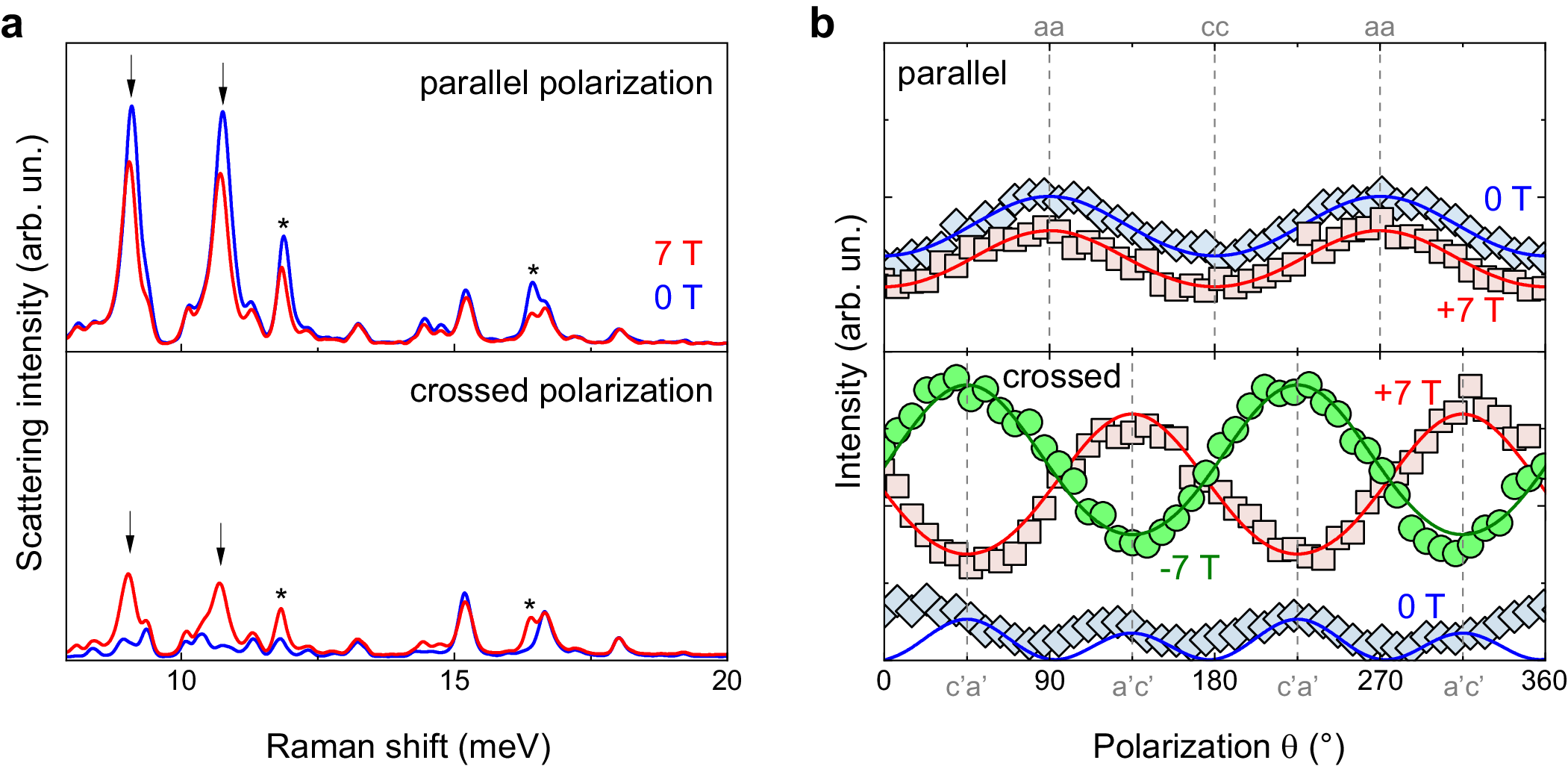}
\caption{\label{fig:2-polar-plots} \textbf{Field-tuning axial Higgs modes.} \textbf{a}, As-measured Raman spectra of GdTe$_3$ collected at $T = 2$ K in parallel (top panel) and crossed (bottom panel) polarization at 0 T and at +7 T. Arrows mark the Higgs-type amplitude modes of the CDW phase. The asterisks mark zone-folded phonons that couple to the CDW. \textbf{b}, Integrated intensity of the amplitude mode at 9.1 meV as a function of polarization direction within the $ac$-plane at various fields in parallel (top panel) and crossed (bottom panel) polarization. Dashed grey lines denote the polarization direction with respect to the crystallographic axes.}
\end{figure*}

Let us first focus on Raman spectra taken at zero applied field in parallel and crossed polarizations, shown as solid blue lines in Fig. 2a top and bottom panels, respectively. The spectra consist of phonons (mainly in the energy range from 12 -- 20 meV) and of CDW amplitude (Higgs-) modes at 9.1 meV and 10.8 meV, marked by black arrows (for a detailed thermal evolution of Raman-active modes see Supplementary Note 1 and Supplementary Fig. 1). In parallel polarization the CDW modes clearly dominate the spectrum, while in crossed polarization their intensities are significantly reduced compared to other phonon modes.

Next we apply an out-of-plane magnetic field to the sample, the corresponding Raman spectra are plotted as solid red lines. In crossed polarization, the CDW amplitude modes, barely observable at 0 T, suddenly dominate the Raman spectrum at 7 T. On the other hand, phonon modes around 13, 15, and 18 meV are robust against magnetic fields. Note that the intensities of two of the phonons (marked by asterisks) mimic the field-induced behavior of the CDW mode. This does not necessarily indicate a field-induced structural transition or lattice distortion, and is most likely related to a coupling between this particular phonon and the CDW (see Supplementary Notes 1 and 2, and Supplementary Fig. 2). As we discuss below, there is a distinct difference in the field evolution between these somewhat anomalous phonons and the CDW excitation. In contrast to the remarkable field-enhancement of CDW modes observed in crossed polarization, we find that in parallel polarization a magnetic field of 7 T only moderately changes the intensities of several excitations, but no fundamental difference is observed between measurements at 0 T and +7 T. To remove any effect related to a particular sample orientation, the spectra shown in Fig. 2a were averaged while rotating the light polarization from 0$^{\circ}$ to 360$^{\circ}$ in either parallel or crossed relation.

A key observation of a previous Raman spectroscopy study on GdTe$_3$ carried out at zero magnetic field was the unusually low two-fold symmetry of the Higgs-type CDW amplitude modes observed in both parallel and crossed polarizations, which was taken as experimental evidence for the simultaneous and spontaneous breaking of multiple symmetries, enabling the formation of a CDW amplitude mode with vector character, i.e., an axial Higgs mode instead of a conventional scalar one~\cite{wang-22}. The symmetry of excitations can be investigated via polarization-resolved Raman spectroscopy by rotating the polarization of the laser light within the crystallographic $ac$-plane from 0$^{\circ}$ to 360$^{\circ}$ (see Supplementary Notes 3 and 4 for the full data set). Extracting the integrated intensity of the amplitude mode at 9.1 meV as a function of polarization and magnetic field yields the plots shown in Fig. 2b (the neighboring mode at 10.8 meV mimics this behavior, as shown in Supplementary Figures 3 and 4). In the upper panel we detail the symmetry of the Higgs mode probed in parallel polarization at 0 T and at +7 T. A clear two-fold (180$^{\circ}$) symmetry is observed with and without magnetic fields, in good agreement with the previous Raman study~\cite{wang-22}. In crossed polarization, shown in the bottom panel, a magnetic field dependence becomes strikingly clear: while at zero magnetic fields the symmetry is close to four-fold (with a weak two-fold modulation on top; see Supplementary Note 5 with Supplementary Figures 5 and 6 for a detailed analysis of the periodicity at $B \to 0$), at applied magnetic fields the two-fold symmetry becomes remarkably dominant. Moreover, a 90$^{\circ}$ phase shift is induced by changing the out-of-plane field direction from positive to negative. In both configurations, parallel and crossed, the intensity and periodicity of phonon modes that do not directly couple to the CDW remain fully field-independent (see Supplementary Note 6 with Supplementary Fig. 7 for line cuts of phonons at 15 and 18 meV).

\subsection{\label{symmetry}Implications of the Higgs-mode character on its symmetry}

To account for such low two-fold symmetry in both parallel and crossed scattering configurations requires a Raman tensor with anti-symmetric off-diagonal tensor elements, as described in a previous Raman scattering study~\cite{wang-22}:

\begin{center}
\mbox{$R_{\mathrm{axial}}$=$\begin{pmatrix} \alpha & ... & \beta\\ ... & ... & ...\\ -\beta & ... & \gamma\\
\end{pmatrix}$}
\end{center}

Here, the empty tensor elements (...) reflect the notion that we carry out experiments exclusively within the $ac$ plane, and therefore cannot access any tensor elements related to the crystallographic $b$ direction. In the parallel configuration, the polarizations of incident and scattered light are parallel to each other, \textbf{e}$_{\mathrm{in}}$ // \textbf{e}$_{\mathrm{out}}$, while in crossed configuration \textbf{e}$_{\mathrm{out}}$ is shifted by 90$^{\circ}$ with respect to \textbf{e}$_{\mathrm{in}}$. As the Raman scattered intensity $I$ is given by $I \propto \vert$ \textbf{e}$_{\mathrm{in}} \cdot R_{\mathrm{axial}} \cdot$ \textbf{e}$_{\mathrm{out}} \vert ^2$, we can express the angular dependence of the Higgs mode intensity as $I(\theta) \propto \vert \gamma+(\alpha-\gamma)\cdot \mathrm{cos}^2\theta \vert ^2$ for parallel polarization, and $I(\theta) \propto \vert 1/2 \cdot (\alpha-\gamma) \cdot \mathrm{sin}(2\theta)-\beta \vert ^2$ for crossed polarization. As we note from these equations, the Raman scattering intensity in parallel polarization is solely determined by the diagonal tensor elements $\alpha$ and $\gamma$, while off-diagonal element $\beta$ becomes relevant for the scattering intensity in crossed polarization. Thus, using the dataset recorded in parallel polarization we extract values for $\alpha$ and $\gamma$ (shown in Supplementary Note 7), which then allow us to accurately determine $\beta$ as a single fitting parameter from the crossed-polarization dataset. Fits of these angle-dependent intensity curves to the data shown in Fig. 2b are indicated by solid lines. Based on these fits and on the stark difference between parallel and crossed configurations with and without applied fields, we conclude that the tensor elements $\alpha$ and $\gamma$ remain (mostly) field-independent, while the field-dependent Higgs mode intensity observed in crossed polarization is dictated by the field-dependence of Raman tensor element $\beta$. We also recall that -- based on symmetry considerations -- effects of hybridization between $p_x$ and $p_y$ should only be accessible in crossed polarization and absent in parallel configuration~\cite{eiter-13}. Therefore, tensor element $\beta$ may offer a direct glimpse into the effects of applied magnetic field on orbital hybridization.

\begin{figure*}
\includegraphics[width=14cm]{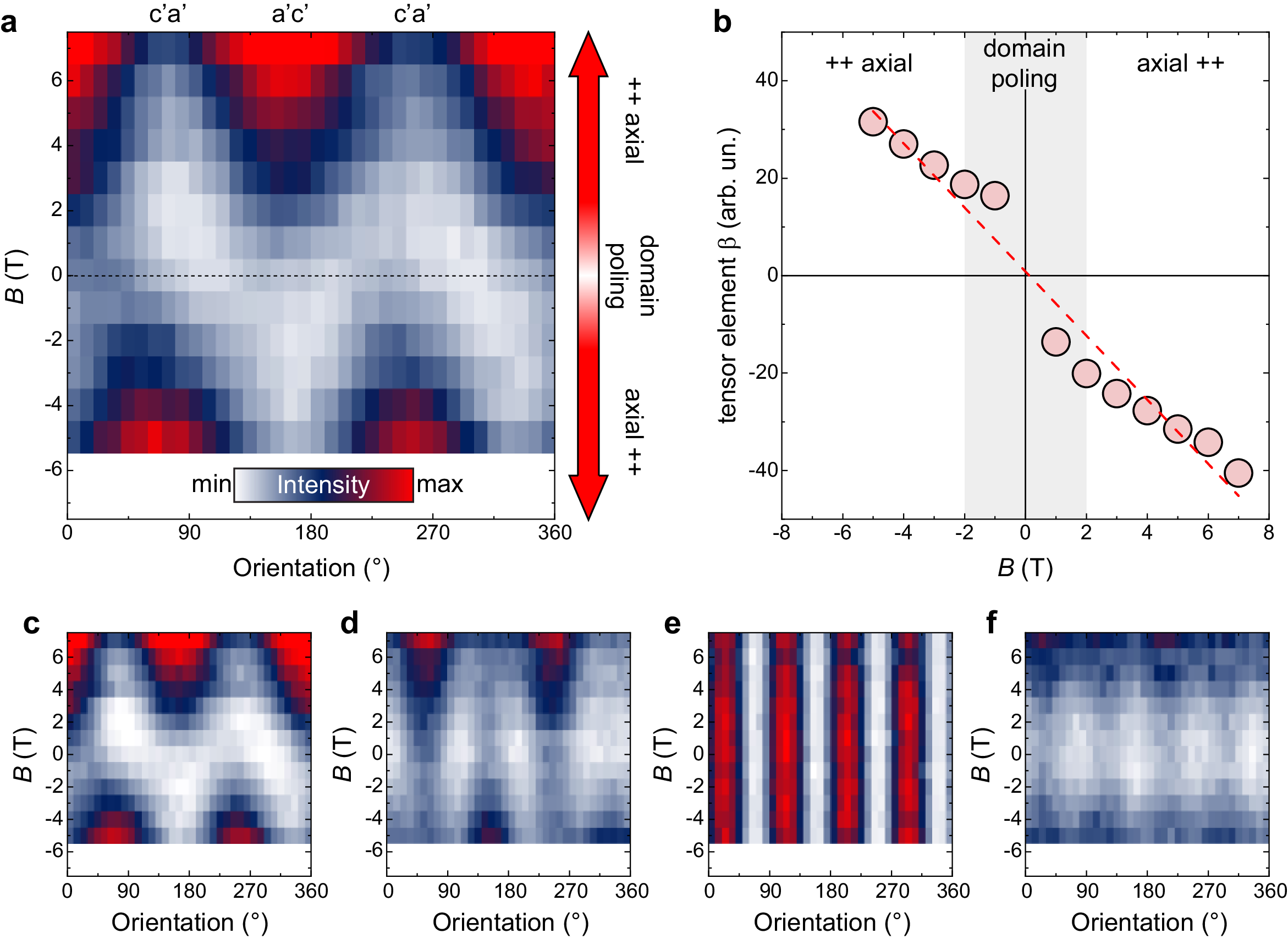}
\caption{\label{fig:3-tensor-element} \textbf{Field-evolution of off-diagonal Raman tensor elements.} \textbf{a}, Integrated scattering intensity of the CDW amplitude mode at 9.1 meV in GdTe$_3$ as a function of out-of-plane field strength and light polarization with \textbf{e}$_{\mathrm{in}} \perp$ \textbf{e}$_{\mathrm{out}}$. At small fields domain poling effects contribute, while at higher fields the axialness of the Higgs mode increases linearly with B, as indicated by the arrow on the right. Crystallographic axes are indicated at the top. All data was acquired at $T = 2$ K. \textbf{b}, Field-dependent Raman tensor element $\beta$ extracted from fits to the integrated intensity plotted in panel a. The standard deviation for each fitted value is within the size of the symbols. The data at $B=0$ has been omitted due to ambiguity in fitting (see Supplementary Note 7 and Supplementary Figures 8-11 for details). The dashed line is a guide to the eyes. \textbf{c-f}, Field- and polarization angle dependence of the Raman scattering intensity for four selected modes (10.8 meV, 12 meV, 15 meV, and 17 meV) measured with \textbf{e}$_{\mathrm{in}} \perp$ \textbf{e}$_{\mathrm{out}}$ at $T = 2$ K.}
\end{figure*}

To quantify the field-dependence of $\beta$, we now continuously tune the out-of-plane magnetic field from +7 T to -5 T in steps of 1 T, and plot the intensity of the CDW amplitude mode measured in crossed configuration as a function of in-plane light polarization in the color contour plot of Fig. 3a. Note that these detailed field-dependent measurements have been carried out on a different GdTe$_3$ specimen (``sample 2'') from the one discussed in Fig. 2 (``sample 1''). This has been done to demonstrate reproducibility of the observed field-dependent effect. In Fig. 3b we plot the fitted values of the Raman tensor element $\beta$ as a function applied magnetic field (red circles). For direct comparison, we show the extracted values of $\beta$ from ``sample 1'' and from a third specimen ``sample 3'' in Supplementary Fig. 11. For all three samples $\beta$ becomes strongly enhanced with increasing magnetic fields, and the observed phase shift shown in Fig. 2b for reversing magnetic field directions necessitates a sign change of $\beta$ as a function of field direction. Summarizing our results, we find that $R_{\mathrm{axial}}$ can aptly describe our data, and that tensor element $\beta$ appears to be close to proportional to the applied magnetic field $B$ (see Supplementary Note 8 and Supplementary Figures 12 and 13). An important question is whether this field-tuning is uniquely observed for the CDW amplitudon, or if other excitations show comparable anomalies. Indeed, the high-energy CDW shoulder located at 10.8 meV shows an identical behavior, as seen in Fig. 3c. The zone-folded phonon located at 12 meV also evidences some appreciable field-dependence (Fig. 3d), although it is more subtle and with a 90$^{\circ}$ phase shift with respect to the amplitudon. In Fig. 3e we plot the field-dependence of a reference phonon located at 15 meV with no field-dependence whatsoever, while a second zone-folded phonon at 17 meV keeps its four-fold periodicity and phase independent of the magnetic field, and only evidences a general field-induced increase in intensity (Fig. 3f). We thus conclude that only the CDW amplitudon is dramatically affected by magnetic fields, and that some of the CDW-coupled zone-folded phonons show their own distinct but more subtle field-induced anomalies. This also underlines our assumption that all observed effects are intrinsic to the sample rather than extrinsic experimental artifacts, which should not distinguish between amplitudons and phonons. In the following, we will rationalize the observed field stabilization of the axial Higgs mode, the phase shift with reversing field direction, and the existence of off-diagonal anti-symmetric tensor elements in $R_{\mathrm{axial}}$, by considering either the orbitals or ferro-rotational lattice distortion involved in the CDW and Raman scattering processes.

\section{\label{discussion}Discussion}

\begin{figure*}
\includegraphics[width=16cm]{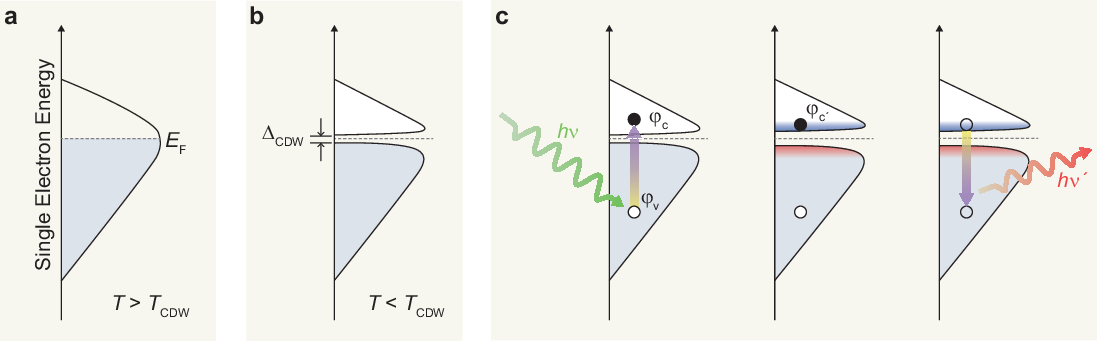}
\caption{\label{fig:4-Higgs-sketch} \textbf{Raman scattering process in the framework of single-particle spectral function involving field-tuned axial Higgs modes.} \textbf{a}, A gapless state at $T > T_{\mathrm{CDW}}$. \textbf{b}, Below $T_{\mathrm{CDW}}$ a charge-density wave gap ($\Delta_{\mathrm{CDW}}$) opens around the region relevant for the Raman scattering process. The system is in its ground state. \textbf{c}, The Raman scattering process probing the axial Higgs mode: An incident photon with energy $h \nu$ (green arrow) excites an electron from the valence band to an unoccupied state across $\Delta_{\mathrm{CDW}}$, which couples to an amplitudon (red and blue shaded areas around $\Delta_{\mathrm{CDW}}$) in a second step. In the final step, the electron recombines with the hole and emits a photon with energy $h \nu'$ (orange arrow) with a polarization orthogonal to that of the incident photon.}
\end{figure*}

The drastic effect of out-of-plane fields on the Higgs mode intensity and on its symmetry requires the presence of intrinsic magnetic degrees of freedom in GdTe$_3$. Here, we will propose and discuss three scenarios, following the illustrations shown in Fig. 1a. Considering the magnetic field dependence, a plausible scenario is that spin degrees of freedom in GdTe$_3$ are the relevant ingredient. In GdTe$_3$, the Gd spins order antiferromagnetically below $T_{\mathrm{N}} = 11.5$ K~\cite{iyeiri-03} with magnetic moments aligned within the plane, which could result in an interplay between magnetic order and the CDW. We can rule this first scenario as unlikely based on the following observations: As shown in Supplementary Notes 9 and 10 with Supplementary Figures 14 and 15, the same field-induced behavior of the Higgs modes is seen below and well above $T_{\mathrm{N}}$. Furthermore, a detailed look at the thermal evolution of Raman active modes reveals the absence of any significant anomalies across $T_{\mathrm{N}}$. More importantly, the non-magnetic sister compound LaTe$_3$ evidences an identical response of its CDW mode to applied magnetic fields.

A second, more likely scenario that is discussed in the context of order parameters with axial vector character is ferro-rotational order~\cite{cheong-18, hlinka-16, liu-23}. Such order, driven by lattice distortion within the CDW phase, does not by itself couple to electromagnetic fields. It may however activate chiral phonons that in turn generate intrinsic magnetic moments~\cite{ataei-24}, or affect the orbital overlap, which could be further tuned by magnetic fields. Our data may, to some extent, support this scenario, as two $A_g$-symmetric phonon modes respond to applied magnetic fields (see the asterisk-marked phonons in Fig. 2a). On the other hand, if the drastic amplitude-mode field-tuning is a secondary effect of magneto-elastic coupling, one might expect a significant continuous evolution of phonon frequencies with magnetic fields. In that sense none of the Raman-active phonons suggest any obvious magneto-elastic coupling (Supplementary Note 2), and clear fingerprints for chiral phonons are lacking (Supplementary Note 11 and Supplementary Fig. 16).

Our third scenario is based on a direct orbital-driven mechanism without invoking lattice degrees of freedom and involves field-tuned hybridization between Te $p_x$ and $p_y$ orbitals. As was pointed out in previous studies, the interference between \textbf{q}$_{\mathrm{CDW}}$ and the additional CDW vector \textbf{c}$^*$-\textbf{q}$_{\mathrm{CDW}}$ resulting from such hybridization determines the axialness of the Higgs mode and thereby its Raman scattering intensity for light polarization configurations compatible with the wave vector~\cite{eiter-13, wang-22}. A hybridization $p_x \pm \mathrm{i}p_y$ is associated with a finite OAM \textbf{m}$_{\ell}$. Hence, applying a magnetic field \textbf{B} will directly affect the mixing of orbitals. In the balanced case (\textbf{m}$_{\ell}$=0) the hybridization and the interference with $c^*-q_{\mathrm{CDW}}$ both vanish. Therefore, the amplitudon would recover its conventional scalar form. Once \textbf{m}$_{\ell}$ assumes a finite value (i.e., the degeneracy between $(p_x + \mathrm{i}p_y)$ and $(p_x - \mathrm{i}p_y)$ is lifted upon application of a magnetic field), hybridization becomes finite and therefore an interplay between \textbf{q}$_{\mathrm{CDW}}$ and \textbf{c}$^*$-\textbf{q}$_{\mathrm{CDW}}$ dictates the Raman scattering intensity of the now axial Higgs mode (indicated by the red- and blue-shaded regions around the Fermi energy $E_{\mathrm{F}}$ in Fig. 4). The corresponding transitions involved in this Raman scattering process are sketched in Fig. 4 in the framework of a single-particle spectral function. Although CDW order leaves the Fermi surface partially ungapped~\cite{ru-08}, we here consider the region in momentum space that is most relevant for our Raman scattering process. Based on this scenario the off-diagonal Raman tensor element can be expressed as $\beta ~ \sim \langle \varphi_{\mathrm{v}} \vert \hat{y} \vert \varphi_{\mathrm{c'}} \rangle \langle \varphi_{\mathrm{c'}} \vert \hat{H}_{\mathrm{el-CDW}} \vert \varphi_{\mathrm{c}} \rangle \langle \varphi_{\mathrm{c}} \vert \hat{x} \vert \varphi_{\mathrm{v}} \rangle$, where $\hat{x}$ and $\hat{y}$ are orthogonal electric dipole operators~\cite{cardona-75, pimenta-21}, and $\varphi_{\mathrm{v}}$ and $\varphi_{\mathrm{c}}$ ($\varphi_{\mathrm{c'}}$) are electronic states in the valence- and conduction band. Here, the Hamiltonian $\hat{H}_{\mathrm{el-CDW}}$ yields the off-diagonal hopping element $\hat{H}_{\mathrm{xy}}$, when $\varphi_{\mathrm{c}}$ and $\varphi_{\mathrm{c'}}$ are orthogonal to each other. In this scattering configuration, $\beta$ is intimately linked to $\hat{H}_{\mathrm{xy}}$, and thereby allows us to directly probe the degree of hybridization between $p_x$ and $p_y$, which is governed by $\hat{H}_{\mathrm{xy}} \sim$ \textbf{B}$\cdot$\textbf{m}$_{\ell}$. This also implies that, in contrast to out-of-plane magnetic fields, in-plane magnetic fields would leave the hybridization between $p_x$ and $p_y$ invariant, resulting in a field-independent Raman response of the CDW amplitudon. Future magneto-Raman scattering experiments carried out in Voigt geometry may therefore ultimately confirm our hypothesis.

Finally, let us comment on subtle differences between our work and the previous Raman scattering study on GdTe$_3$. While Wang, et al., reported a pronounced two-fold symmetric (i.e., axial) Higgs mode in crossed polarization for $B=0$ at both room temperature as well as 8 K~\cite{wang-22}, our data taken at 2 K without magnetic field is more ambiguous with a weak and almost 4-fold periodic Higgs mode. A key difference between our methodologies is the laser spot diameter during the Raman experiment, which is as tight as 2 $\mu$m in the former case, while in our experiment it is spread out to about 100 $\mu$m, thereby averaging over a much larger sample area. To facilitate a better comparison, we also conducted room temperature experiments at zero fields with a tightly focused spot diameter of about 2 $\mu$m, which yielded a clear two-fold periodicity (see Supplementary Note 5). Based on these observations, we conjecture that twin domains are prevalent throughout the sample on a length scale of a few micrometers, which will mostly average out any intrinsic imbalance between +\textbf{m}$_{\ell}$ and -\textbf{m}$_{\ell}$ at $B=0$ T when integrating the Raman signal over a large-enough sample area. Indeed, such twin domains were recently uncovered in GdTe$_3$ via scanning tunneling microscopy~\cite{lee-23}, and their existence would naturally explain the subtle hysteresis behavior seen around small fields for $\beta(B)$ in Fig. 3(b). In Supplementary Note 12 and Supplementary Fig. 17 we compare room temperature spectra of GdTe$_3$ with and without magnetic field, which clearly highlight a field-induced increase in Higgs mode intensity. Conspicuously, all these observations demand that a weaker axial character of the Higgs mode is already developed in GdTe$_3$ at room temperature and without any applied magnetic field. Yet, the continuous linear increase of $\beta$ for higher fields supports the scenario of a field-driven control over axialness. An important implication is that with the onset of CDW order either ferro-rotational distortion spontaneosly appears or time-reversal symmetry breaks spontaneously, even at $B = 0$ T. To test this hypothesis, we call on future $\mu$SR investigations, circular-dichroism angle resolved photoemission spectroscopy, or scanning tunneling microscopy experiments in applied magnetic fields. If indeed time-reversal symmetry breaks spontaneously in $R$Te$_3$, such studies could reveal vital insight relevant for related materials with unconventional charge-density wave order, such as vanadium-based kagome metals. Furthermore, these rare-earth tritellurites may be apt hosts for other exotic states of matter, such as chiral superconductivity, e.g., by applying hydrostatic pressure~\cite{zocco-15}.

\section{Methods}

\textbf{Sample synthesis.} Stoichiometric single-crystalline GdTe$_3$ flakes were used for the self-flux growth method using a box furnace. High-purity Gd metal (99.9\%) and Te chips (99.999\%) were mixed in a 1:30 molar ratio to achieve a Te-rich self-flux condition. The mixed precursor of GdTe$_3$ was loaded into quartz tubes and sealed at $\sim 10^{-5}$ torr using a turbo pump to prevent oxygen contamination. The maximum heating temperature was set to $900^{\circ}$C, and the temperature was maintained for 24 hours to achieve a homogeneous melt. After melting, the sample was cooled down to $500^{\circ}$C at a rate of $-2^{\circ}$C per hour. The melt in the ampule was decanted at room temperature until it turns solids. 

\textbf{Raman scattering.} Samples were mechanically exfoliated right before being transferred into the He-gas filled sample chamber of a magneto-optical cryostat (Oxford SpectromagPT, $T_{\mathrm{min}} = 1.6$ K, $B_{\mathrm{max}} = \pm 7$ T). Thereby, the fresh surface exposure to air was minimized to less than 5 seconds. While the sample remained inside the cryostat no effects of sample degradation through the appearance of additional tellurium oxide modes~\cite{gray-20} were observed over a period of two weeks. Raman scattering experiments were carried out in backscattering geometry using a single-mode laser emitting at $\lambda = 561$ nm (Oxxius-LCX) and a laser power of 0.6 mW or less at the sample position. The laser was focused onto the sample via a series of achromatic lenses, resulting in a beam spot diameter of about 100 $\mu$m (for details about the beampath see Supplementary Note 13 and Supplementary Fig. 18). The in-plane light polarization was controlled using a superachromatic $\lambda$/2 waveplate (Thorlabs) in front of the sample. Raman-scattered light was dispersed and recorded through a Princeton Instruments TriVista 777 spectrometer and a PyLoN eXcelon charge-coupled device, respectively. For field-dependent measurements samples were first zero-field cooled to $T=2$ K, followed by a magnetic field ramp up to +7 T. From there, data was collected while decreasing the field in steps of 1 T through $B=0$ and down to -5 T. Additional zero-field Raman spectra were collected using a home-built microscope stage with a laser spot diameter at the sample surface of about 2 $\mu$m and the sample mounted inside an open-flow cryostat (Oxford MicroStat HR).

\section{acknowledgments}
We acknowledge important discussions with Ken Burch, Birender Singh, Suyoung Lee, Jennifer Cano, Judy Cha, Rafael Fernandez, and Eun-Gook Moon. This work was supported by the Institute for Basic Science (IBS) (Grant Nos. IBS-R009-G2, IBS-R009-Y3). S.R.P. was supported by the NRF (Grant No. 2020R1A2C1011439).

\section{Author contributions}
J.P. synthesized GdTe$_3$ and LaTe$_3$ single crystals. D.W., J.P., and C.K. performed Raman spectroscopic measurements. D.W. and C.K. analyzed the data. Symmetry analysis was carried out by T.T., S.R.P., and C.K. D.W. and C.K. wrote the manuscript with important contributions from all authors.

\textbf{Supplementary Note 1 $|$ Thermal evolution of Raman-active modes in GdTe$_3$}

\begin{figure*}
\label{figure5}
\centering
\includegraphics[width=10cm]{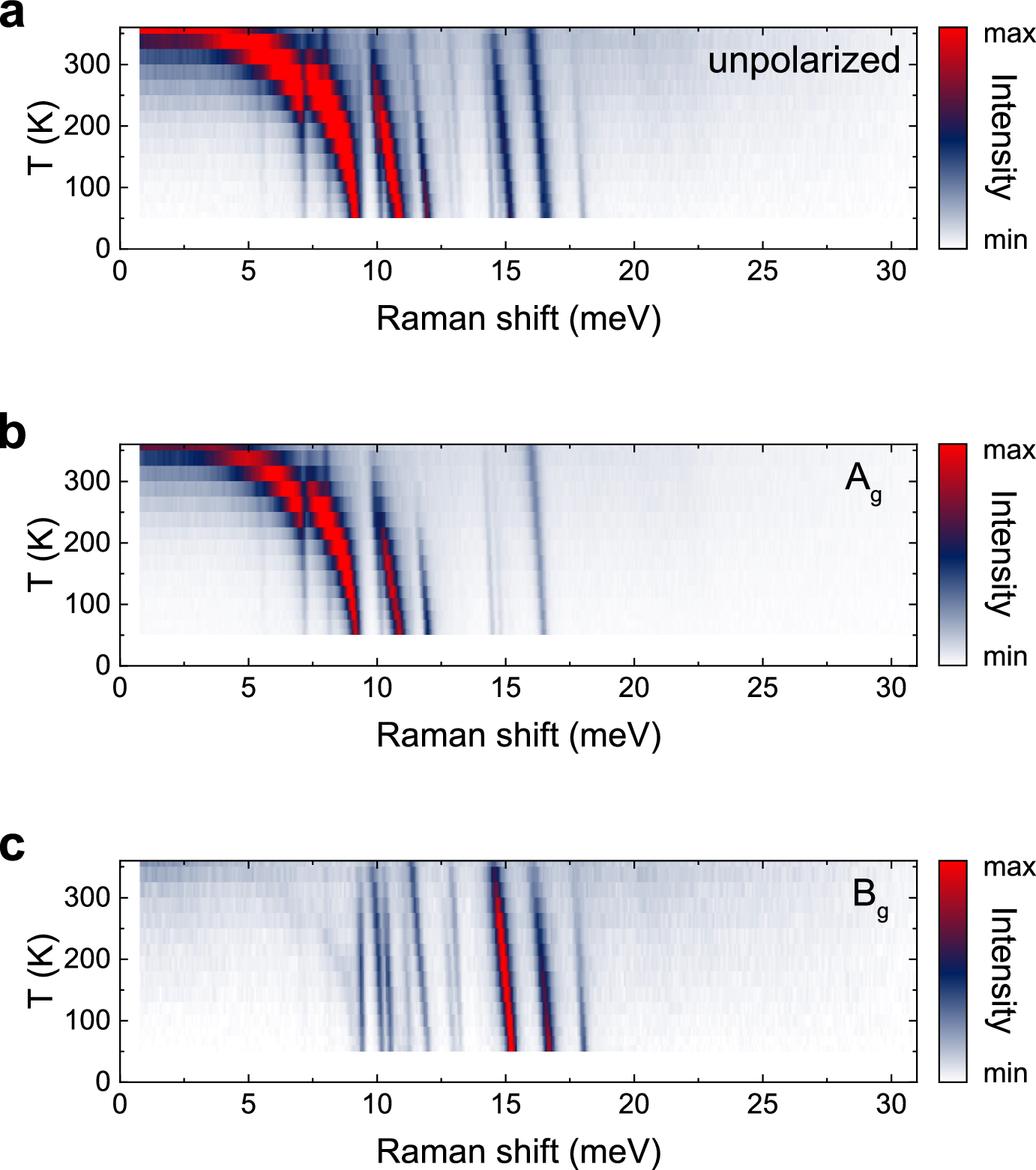}
\caption{\textbf{Suppl. Fig. 1: Detailed temperature-dependence.} Temperature-dependence of Raman-active modes in GdTe$_3$ measured at $B=0$ unpolarized (a), in $A_g$ polarization channel (b), and in $B_g$ polarization channel (c).}
\end{figure*}

In Supplementary Fig. 1 we show color-contour plots of the temperature dependence of Raman-active excitations in GdTe$_3$ in unpolarized condition (panel a). For completeness, we also include the polarization-resolved $A_g$ and $B_g$ channels in panels (b) and (c) to fully trace the temperature dependence of phonons which might (partially) overlap energetically. We find that the amplitudon significantly hardens from about 5 meV at room temperature to 9.1 meV at low temperatures, with a secondary shoulder emerging around 10.8 meV. At 12 meV and at 17 meV, excitations of $A_g$ and of $B_g$ symmetry overlap.

\textbf{Supplementary Note 2 $|$ Field-enhanced Higgs modes}

\begin{figure*}
\label{figure6}
\centering
\includegraphics[width=10cm]{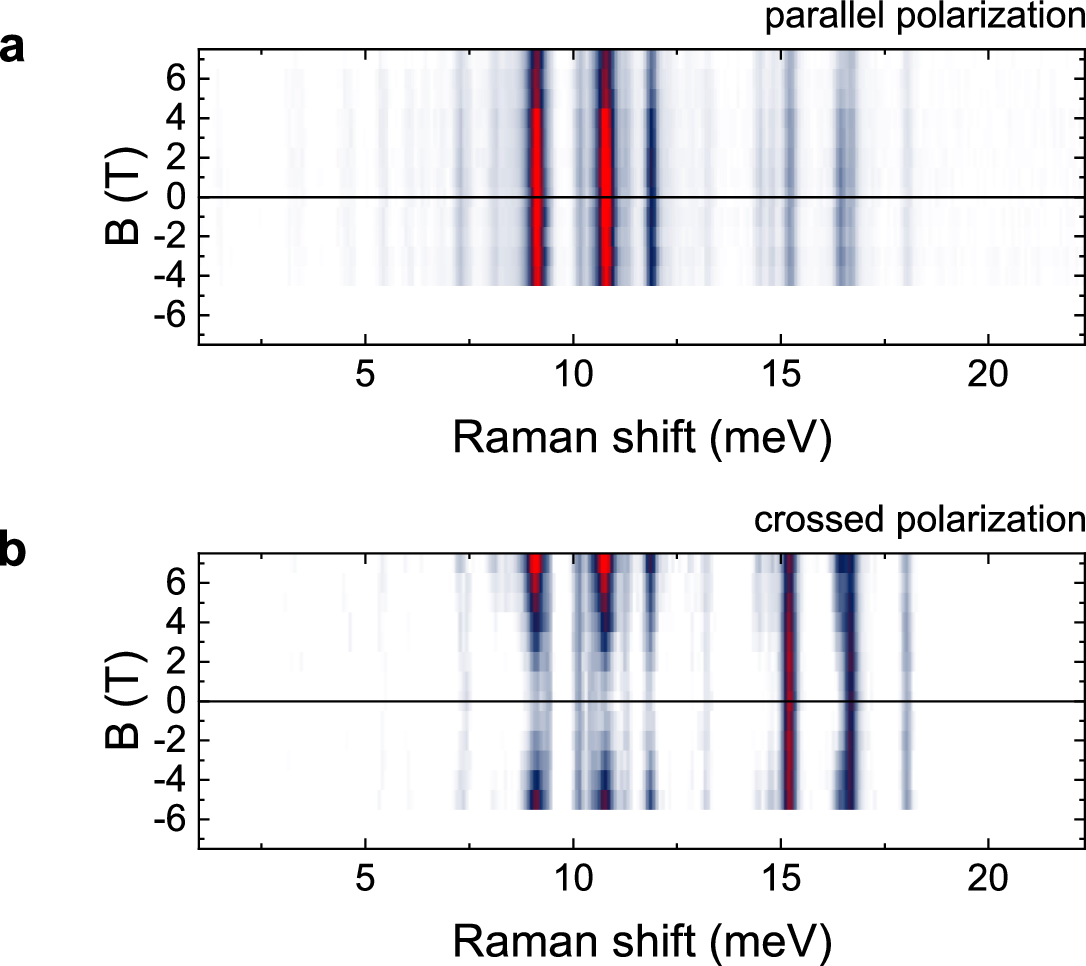}
\caption{\textbf{Suppl. Fig. 2: Detailed field-dependence.} Raman spectra of GdTe$_3$ recorded in parallel (a) and in crossed (b) polarization at $T = 2$ K with out-of-plane magnetic fields ranging from -5 T to +7 T.}
\end{figure*}

In Supplementary Fig. 2 we show Raman spectra obtained in parallel (a) and in crossed (b) polarization with increasing magnetic fields at $T = 2$ K. The plotted spectra are averaged over in-plane polarization angles from $0^{\circ}$ -- $360^{\circ}$. No shift in energy is observed as a function of magnetic field for any of the modes. While the phonon intensities remain mostly unaffected by applied fields and no significant field evolution is evident in parallel polarization, the charge-density-wave modes around 10 meV in crossed polarization strongly gain spectral weight with increasing magnetic fields. Note that one phonon around 16.5 meV in crossed polarization mimics the intensity behavior of the CDW modes, which can be taken as evidence for electron-phonon coupling between these excitations.

\textbf{Supplementary Note 3 $|$ Full dataset recorded in crossed polarization}

\begin{figure*}
\label{figure7}
\centering
\includegraphics[width=10cm]{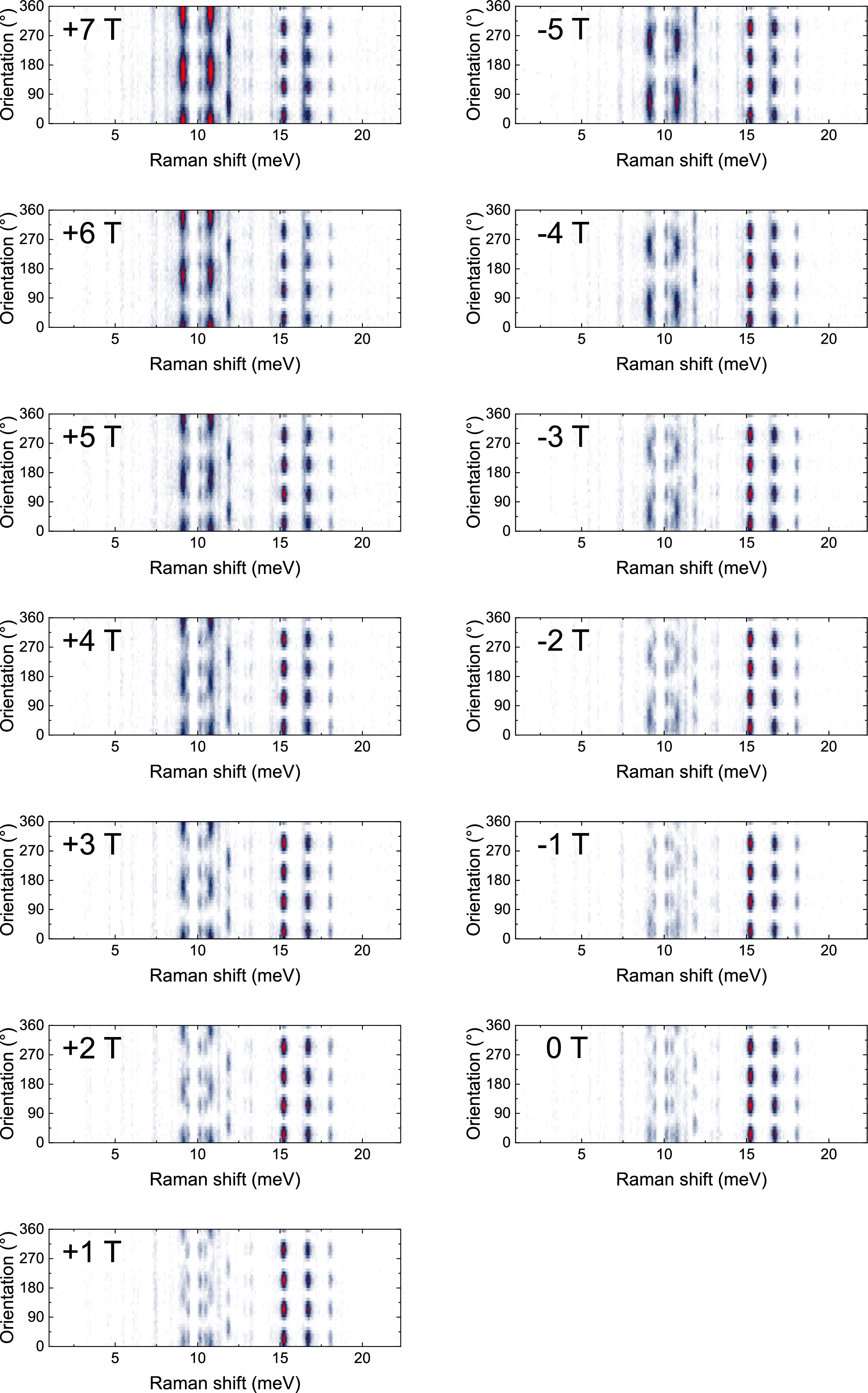}
\caption{\textbf{Suppl. Fig. 3: Field-dependent polarization-resolved Raman plots in crossed configuration.} All data was recorded at $T = 2$ K.}
\end{figure*}

The full dataset of Raman spectra recorded at various applied out-of-plane magnetic fields in crossed polarization on GdTe$_3$ at $T = 2$ K is shown in Supplementary Fig. 3. The amplitude modes are centered around 10 meV and show their distinct field-dependence in terms of intensity and periodicity.

\textbf{Supplementary Note 4 $|$ Full dataset recorded in parallel polarization}

\begin{figure*}
\label{figure8}
\centering
\includegraphics[width=10cm]{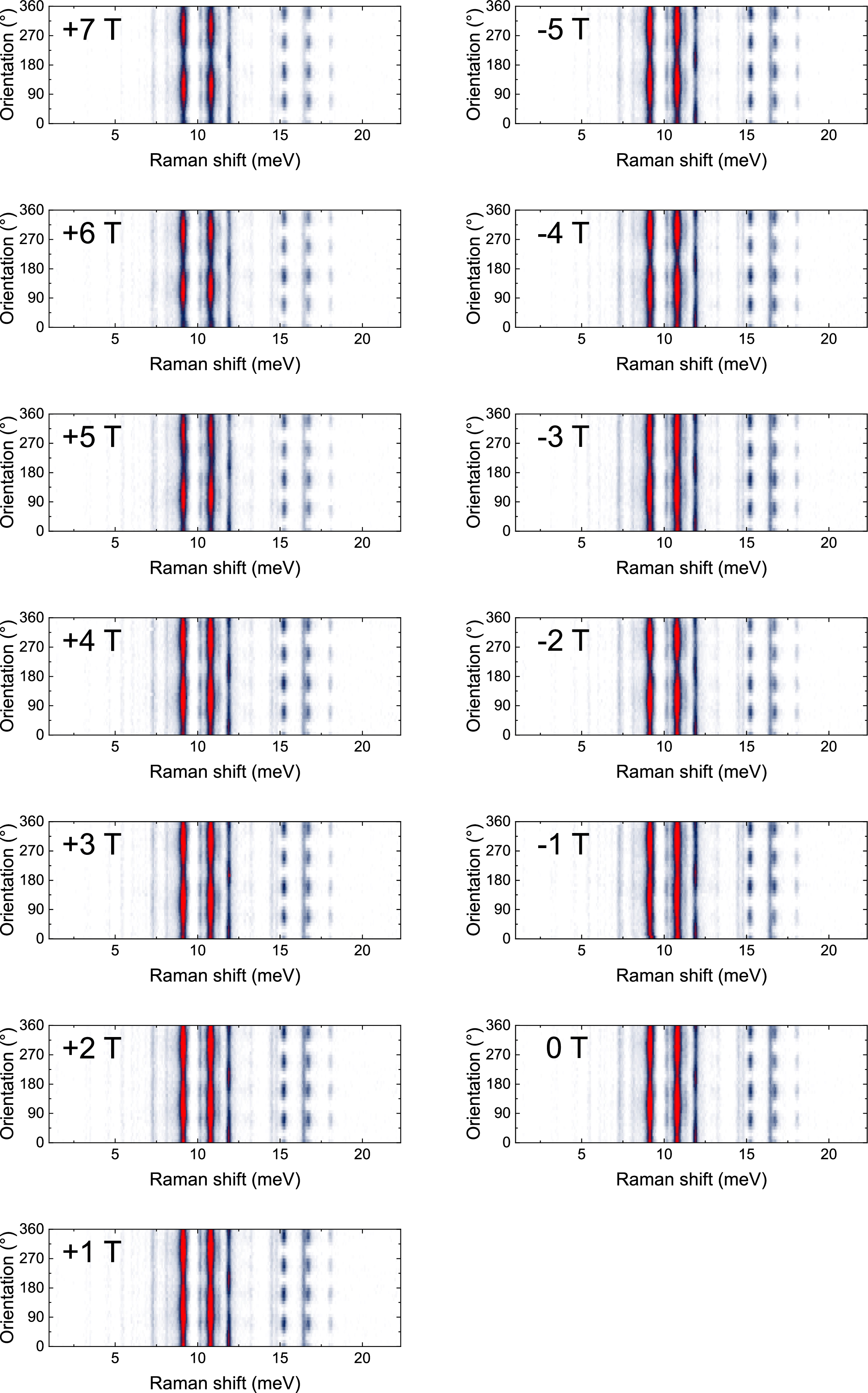}
\caption{\textbf{Suppl. Fig. 4: Field-dependent polarization-resolved Raman plots in parallel configuration.} All data was recorded at $T = 2$ K.}
\end{figure*}

The full dataset of Raman spectra recorded at various applied out-of-plane magnetic fields in parallel polarization on GdTe$_3$ at $T = 2$ K is shown in Supplementary Fig. 4. The amplitude modes are centered around 10 meV and reveal a mostly field-independent behavior.

\textbf{Supplementary Note 5 $|$ Higgs modes in GdTe$_3$ at $B = 0$}

\begin{figure*}
\label{figure9}
\centering
\includegraphics[width=8cm]{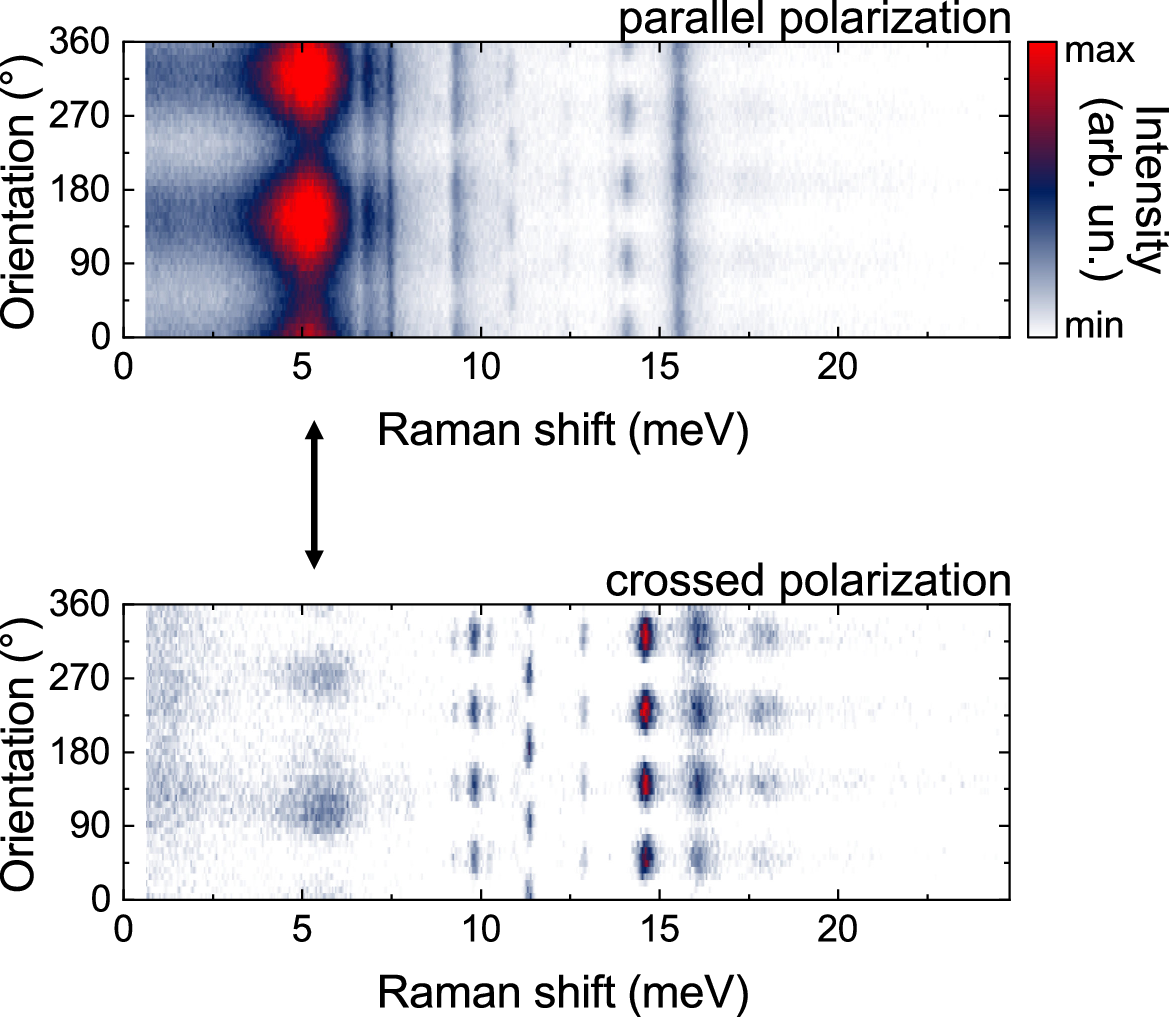}
\caption{\textbf{Suppl. Fig. 5: Polarization-resolved Raman spectra of GdTe$_3$ measured at $T = 300$ K and without applied magnetic field.}}
\end{figure*}

Supplementary Fig. 5 plots Raman spectra recorded at room temperature and without applied magnetic field in parallel and in crossed configuration. These data were taken using a 40$\times$ microscope objective with the sample mounted inside a microscope-type Oxford MicroStat cryostat, allowing for a tight laser spot diameter of about 2 $\mu$m on the sample surface. In this way we closely replicate the experimental conditions used in Ref.~\cite{wang-22}. In contrast, field-dependent data presented throughout this study were obtained from within an Oxford SpectroMag cryostat and a laser spot diameter of about 100 $\mu$m. In the presence of domains, this difference in probing area can potentially result in different observed periodicities. Indeed, the Higgs mode (marked by a black arrow) is characterized by a clear 2-fold symmetry, which necessitates a finite value for the off-diagonal tensor element $\beta$ at $B = 0$ T and $T \gg T_{\mathrm{N}}$. Conversely, when sampling a larger area (see, e.g., Supplementary Figures 3 and 6) the periodicity of the Higgs mode approaches a 4-fold symmetry, which raises the possibility of micrometer-sized domains~\cite{lee-23}.

\begin{figure*}
\label{figure10}
\centering
\includegraphics[width=10cm]{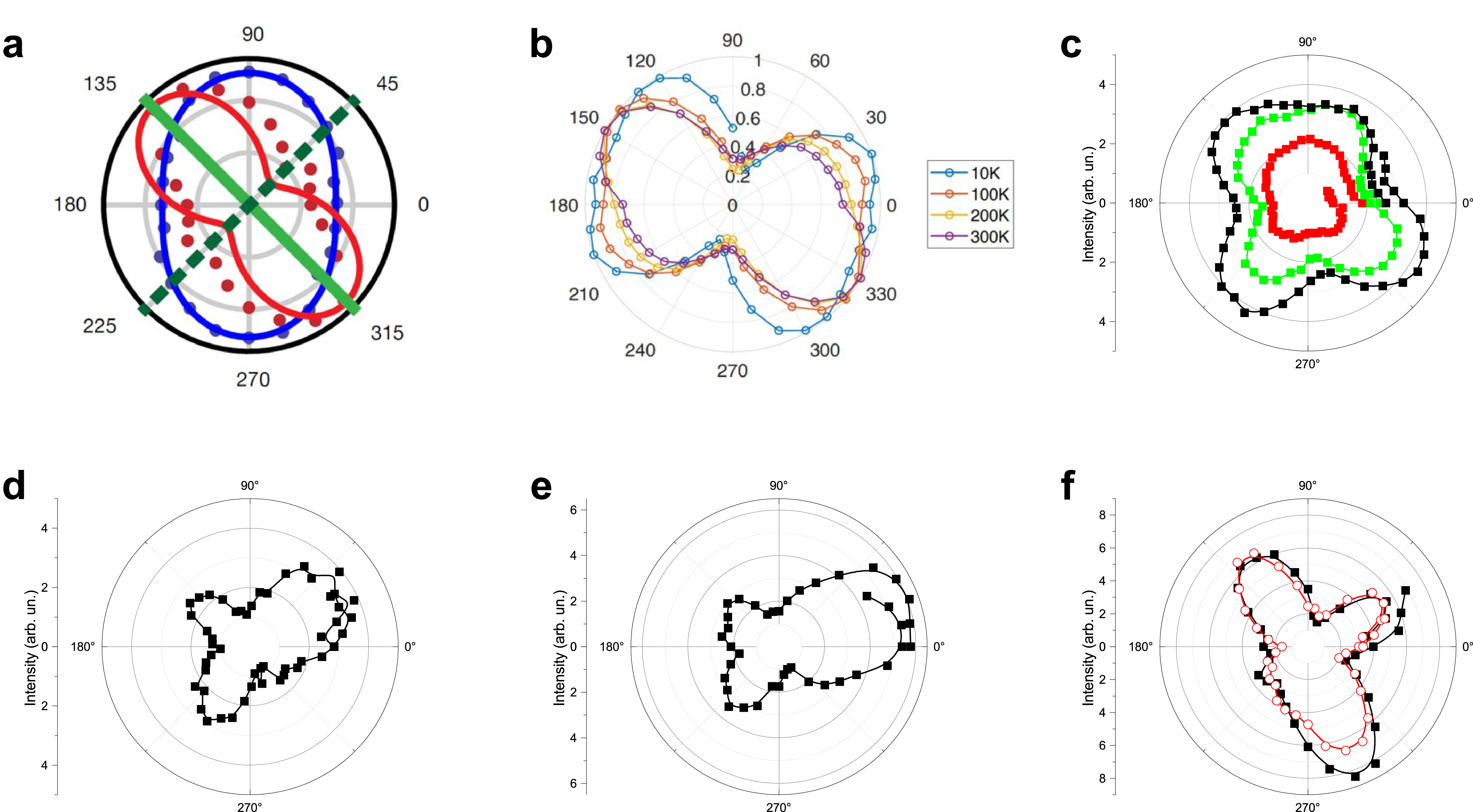}
\caption{\textbf{Suppl. Fig. 6: Polar plots of the Higgs mode intensity.} (a) Red dots: Room temperature data, solid red line: fit. (b) Data taken on a multi-domain flake at various temperatures. Panels (a) and (b) taken from Wang, et al., Nature \textbf{606}, 896. (c) Higgs mode intensity measured at room temperature across a freshly cleaved GdTe$_3$ sample at three different positions (corresponding to black, green, and red data points). Laser spot diameter: 2 $\mu$m. (d)-(f) Higgs mode intensity measured at $T=2$ K and $B=0$ with a spot diameter of 100 $\mu$m on three individual, freshly cleaved GdTe$_3$ samples.}
\end{figure*}

Next, we wish to comment on the ambiguous polarization-dependent periodicity of the Higgs mode in the limit $B \to 0$. Let us start by recalling that small discrepancies between (perfectly two-fold) fit and experiment also exist in some of the polarization-resolved GdTe$_3$-Raman data reported in~\cite{wang-22} (see Supplementary Fig. 6a, which is taken from Fig. 4(c) of their main text). Furthermore, in their Supplementary Fig. 10, reproduced here as Supplementary Fig. 6b, they find that the existence of domains creates a 4-fold pattern with 2-fold distortion, which is in good agreement with our data taken at room temperature across a single GdTe$_3$ flake but measured at three different positions (Supplementary Fig. 6c). In Supplementary Figs. 6d-f we show our data taken inside the magneto-optical cryostat at $T = 2$ K and at $B=0$ on three different, freshly cleaved specimens of GdTe$_3$. For the third sample shown in Supplementary Fig. 6f we repeated the polarization measurements twice to check for consistency. For all three cases the patterns are somewhat distorted with a vestigial two-fold symmetry.

When moving from our room temperature, microscope stage to the magneto-optical cryostat, we increase the spot diameter from about 2 $\mu$m to about 100 $\mu$m. Thereby, domains, domain walls, and disordered regions become more relevant. Although we took great care to align the optical beam path such that the laser spot position on the sample does not move when rotating the polarization, we cannot be absolutely certain that there is no minor movement on the micrometer-scale. In that case we may probe partially different regions and different domain patterns on the sample surface depending on the light polarization. Finally, we also note that at room temperature the CDW mode is energetically separated from any phonons, which may result in more consistent, clean-looking polar plots. Meanwhile, at our base temperature of 2 K, the CDW mode hardens to an energy where it overlaps with many Raman-active phonons (see the temperature-dependent Raman plots in Supplementary Fig. 1). These phonons have their own distinct polarization dependencies and thereby it may be challenging to fully separate the CDW mode from small contributions of neighboring phonons when analyzing the data, especially when the CDW intensity becomes weak in the limit of small fields. Regardless of the exact polarization behavior at $B=0$, we emphasize that our main observation is a dramatic increase in CDW intensity together with its perfect two-fold periodicity in the limit of large fields, and a 90$^{\circ}$ phase shift when reversing the field. These observations can be remarkably well described by considering a field-dependence of the off-diagonal tensor element $\beta$, i.e., $\beta \sim B$.

\textbf{Supplementary Note 6 $|$ Effect of magnetic fields and temperature on phonon symmetries}

\begin{figure*}
\label{figure11}
\centering
\includegraphics[width=10cm]{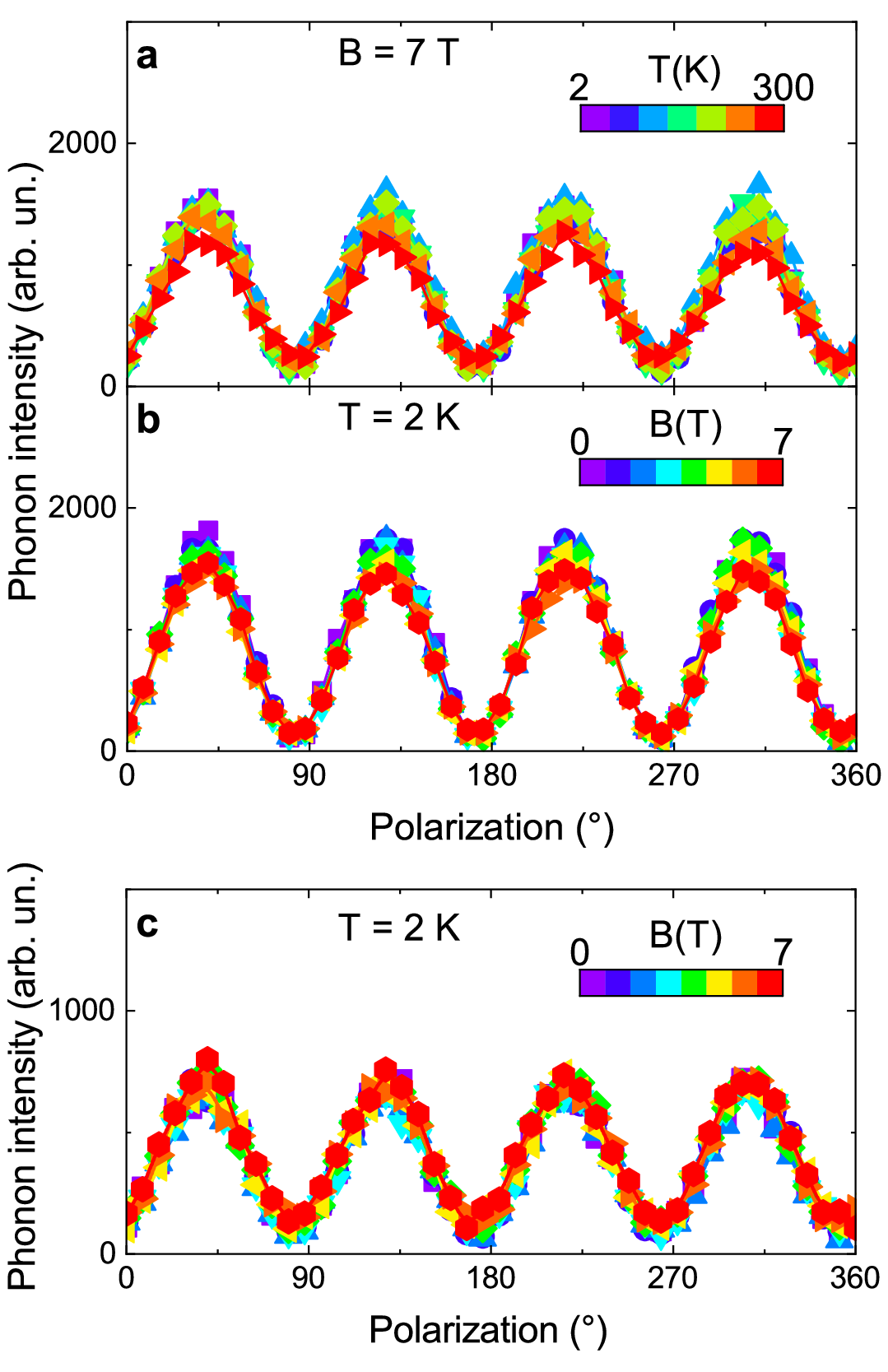}
\caption{\textbf{Suppl. Fig. 7: Effect of magnetic field and temperature on the phonon symmetry.} \textbf{a}, Integrated intensity of the $B_{g}$ phonon at 15 meV measured in crossed configuration at a fixed magnetic field of 7 T at various temperatures 2 K, 50 K, 100 K, 150 K, 200 K, 250 K, and 300 K. \textbf{b}, Integrated intensity of the same phonon as in (a), measured in crossed configuration at a fixed temperature of $T = 2$ K and in applied magnetic fields of 0 T, 1 T, 2 T, 3 T, 4 T, 5 T, 6 T, and 7 T. \textbf{c}, Integrated intensity of the $B_{g}$ phonon at 18 meV measured in crossed configuration at a fixed temperature of $T = 2$ K and in applied magnetic fields of 0 T, 1 T, 2 T, 3 T, 4 T, 5 T, 6 T, and 7 T.}
\end{figure*}

In this Supplementary Note we test the effect of magnetic fields and temperature on two well-defined phonon modes in GdTe$_3$, located at 15 and at 18 meV. In Supplementary Fig. 7a and 7b, polarization-resolved phonon intensity values of the former are shown, which were measured in crossed configuration. Supplementary Fig. 7c plots the field-dependent intensity values of the phonon at 18 meV. In contrast to the Higgs mode, which is strongly affected in intensity and in symmetry by applied magnetic fields, both phonons appear fully unaffected by magnetic field. Therefore, we can also rule out other potential experimental artifacts, such as substantial Faraday rotation of the light polarization through the cryostat window as the source for the anomalous Higgs mode behavior.

\textbf{Supplementary Note 7 $|$ Raman-tensor fits to the amplitude mode in GdTe$_3$}

\begin{figure*}
\label{figure12}
\centering
\includegraphics[width=16cm]{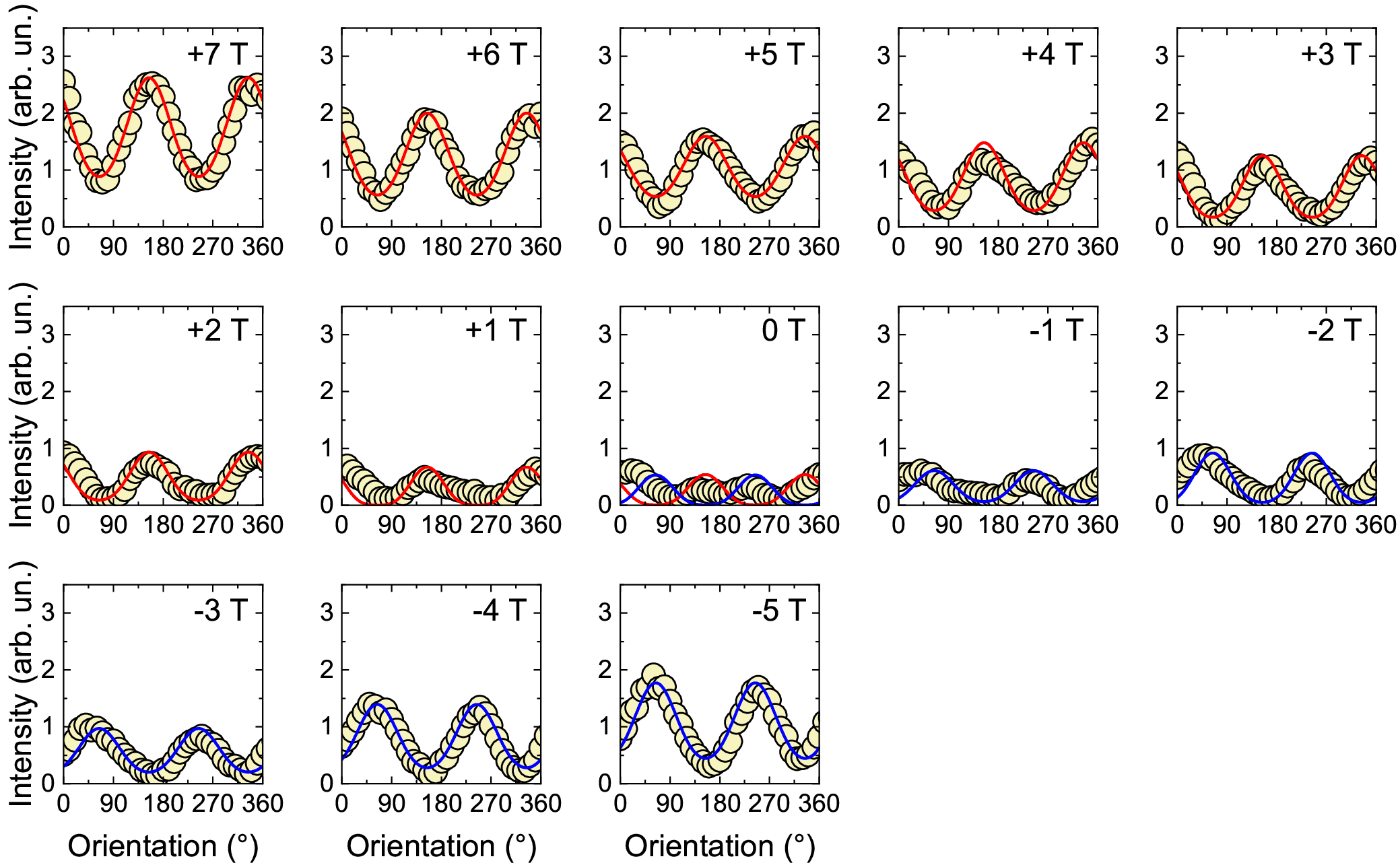}
\caption{\textbf{Suppl. Fig. 8: Full dataset of Raman-tensor fits (solid lines) to the extracted Higgs-mode intensities (filled circles) measured in crossed configuration and at $T = 2$ K as a function of light polarization, magnetic field strength, and magnetic field direction.}}
\end{figure*}

\begin{figure*}
\label{figure13}
\centering
\includegraphics[width=16cm]{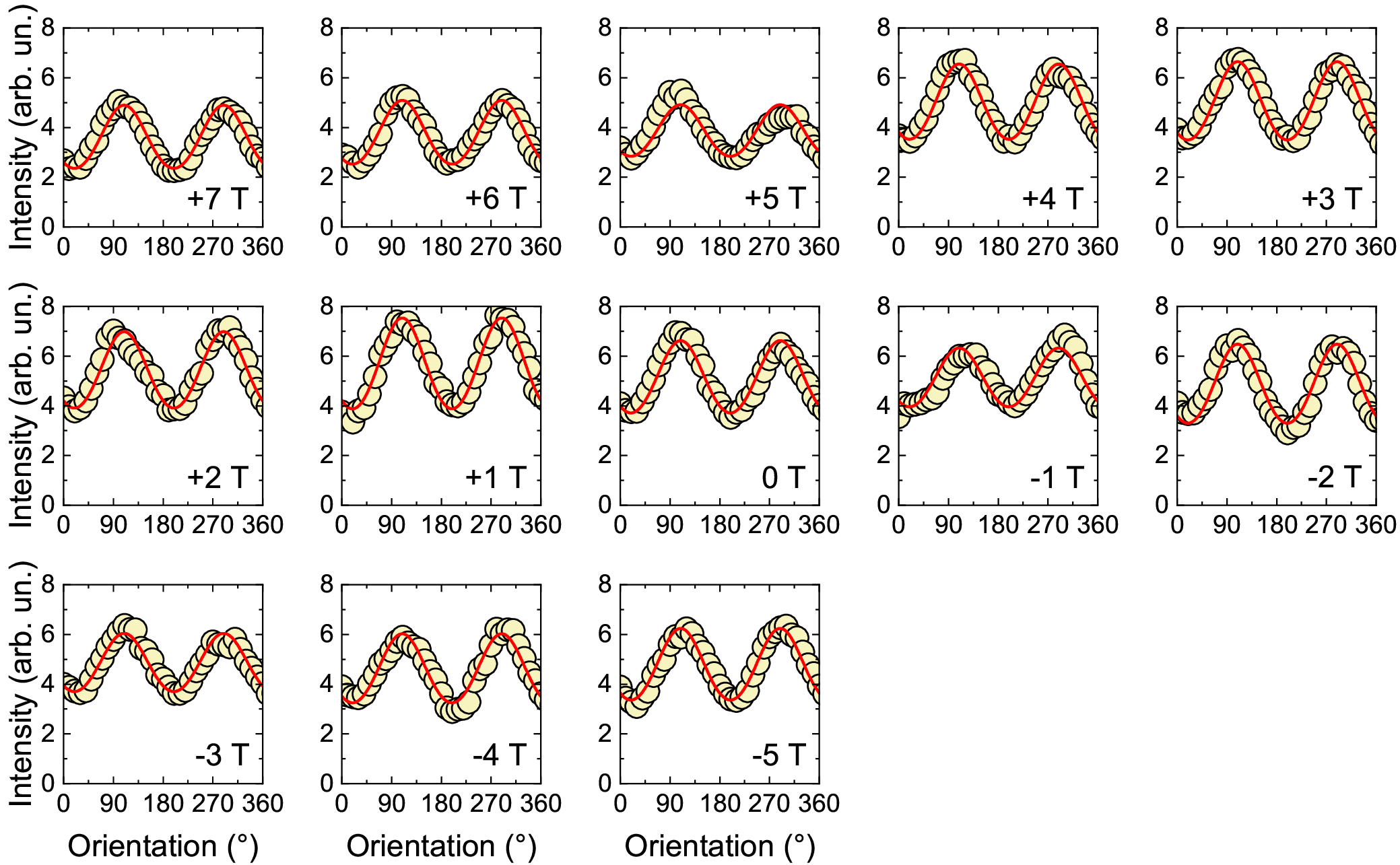}
\caption{\textbf{Suppl. Fig. 9: Full dataset of Raman-tensor fits (solid lines) to the extracted Higgs-mode intensities (filled circles) measured in parallel configuration and at $T = 2$ K as a function of light polarization, magnetic field strength, and magnetic field direction.}}
\end{figure*}

The full dataset of Raman-tensor fits to the Higgs-mode intensity observed in GdTe$_3$ is shown in Supplementary Figs. 8 (for crossed polarization) and 9 (for parallel polarization). In the former case, for fields of $\pm 2$ T and greater, the Raman-tensor yields a very satisfying fit. For weaker magnetic fields of $\pm 1$ T the fits become somewhat ambiguous, which is due to the smaller signal intensity, as well as due to its close vicinity to neighboring phonon modes, which partially overlap and thereby contribute to the extracted intensity. For the case of $B = 0$, the dominating intensity contribution stems from the neighboring phonon modes, and the Higgs mode intensity cannot be extracted unambiguously anymore.
In case of parallel polarization, both periodicity and intensity remain (largely) unaffected, with any changes between fields likely originating from a slight change in sample position. Line cuts of the Higgs mode intensity are shown in a color-contour plot together with extracted tensor element values $\alpha$ and $\gamma$ in Supplementary Fig. 10.

\begin{figure*}
\label{figure14}
\centering
\includegraphics[width=16cm]{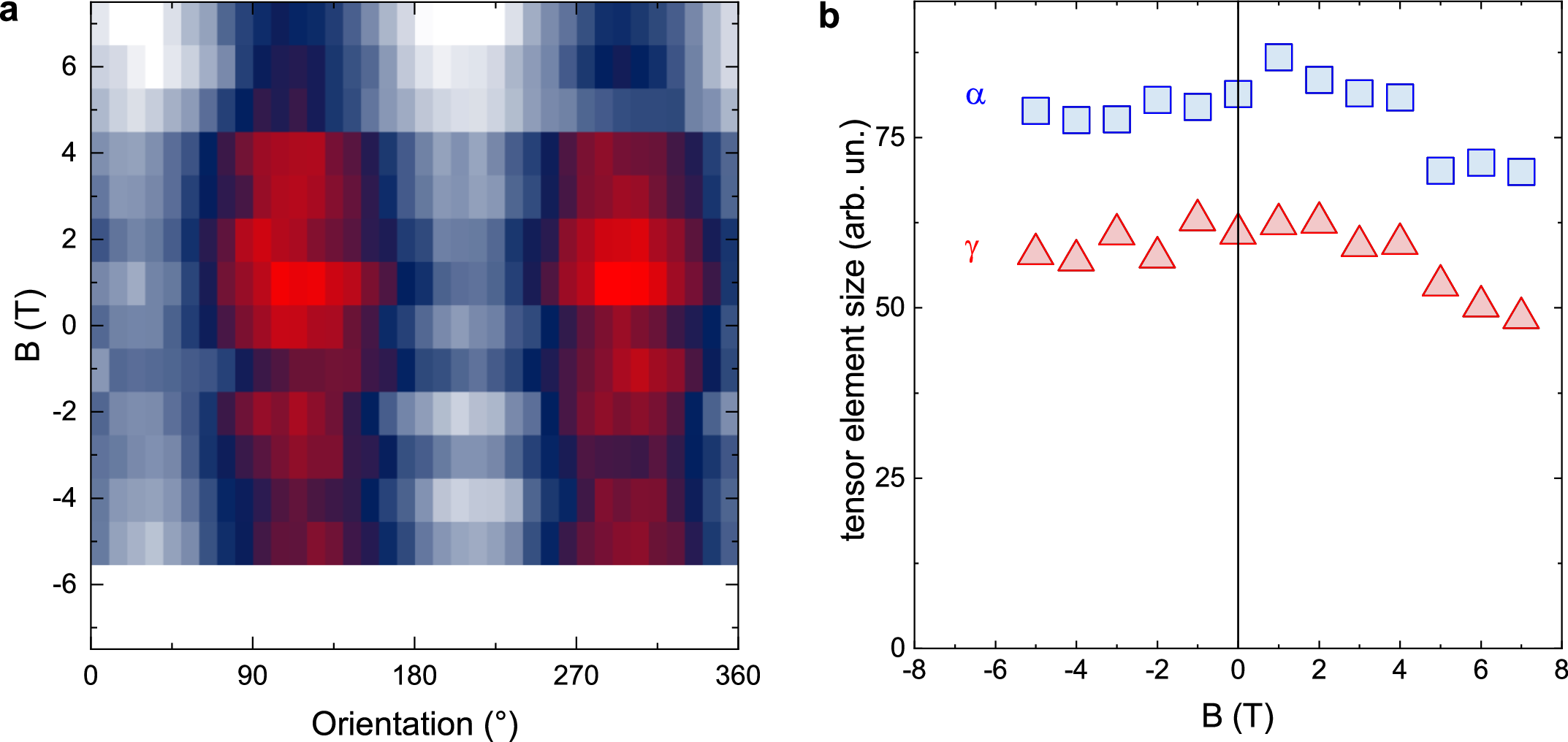}
\caption{\textbf{Suppl. Fig. 10: Higgs-mode field-dependence in parallel polarization.} \textbf{a}, Color-contour plot of the Higgs-mode intensity recorded at $T = 2$ K as a function of polarization angle and magnetic field. \textbf{b}, Fit parameters $\alpha$ and $\gamma$, extracted from the fits shown in Supplementary Fig. 9.}
\end{figure*}

\begin{figure*}
\label{figure15}
\centering
\includegraphics[width=6cm]{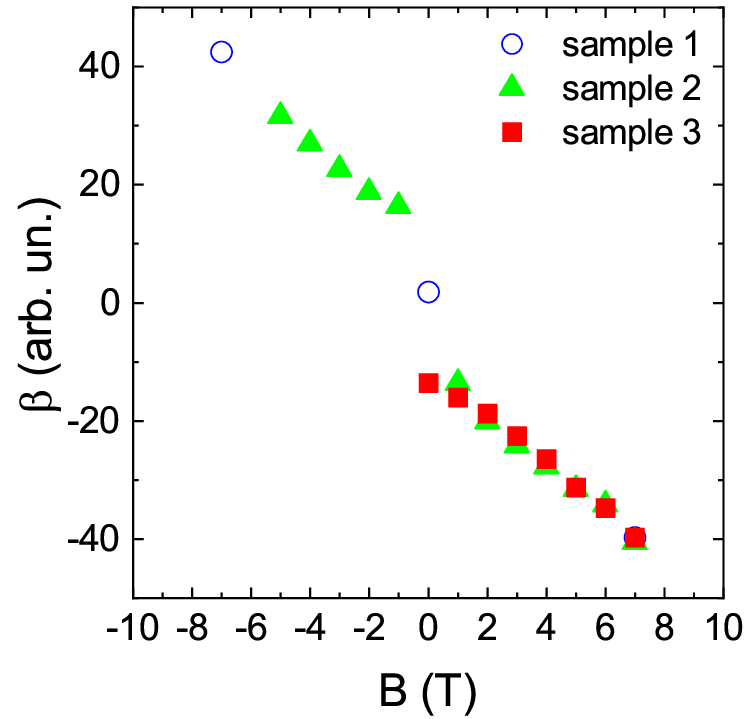}
\caption{\textbf{Suppl. Fig. 11: Tensor element $\beta$ as a function of magnetic field measured in three different GdTe$_3$ samples at $T = 2$ K.}}
\end{figure*}

To test reproducibility of the $\beta \sim B$ scaling (shown in Fig. 3, main text), we plot the field-dependence of the off-diagonal Raman tensor element $\beta$ extracted from three different, freshly exfoliated flakes in Supplementary Fig. 11. We find a remarkable consistency between the three datasets at large fields, with small deviations occurring for small $B$. These deviations can be related to the overall weak signal intensity of the Higgs mode at low fields, or due to domain effects, which become more pronounced as $B \to 0$.

\textbf{Supplementary Note 8 $|$ Raman tensor simulations}

\begin{figure*}
\label{figure16}
\centering
\includegraphics[width=12cm]{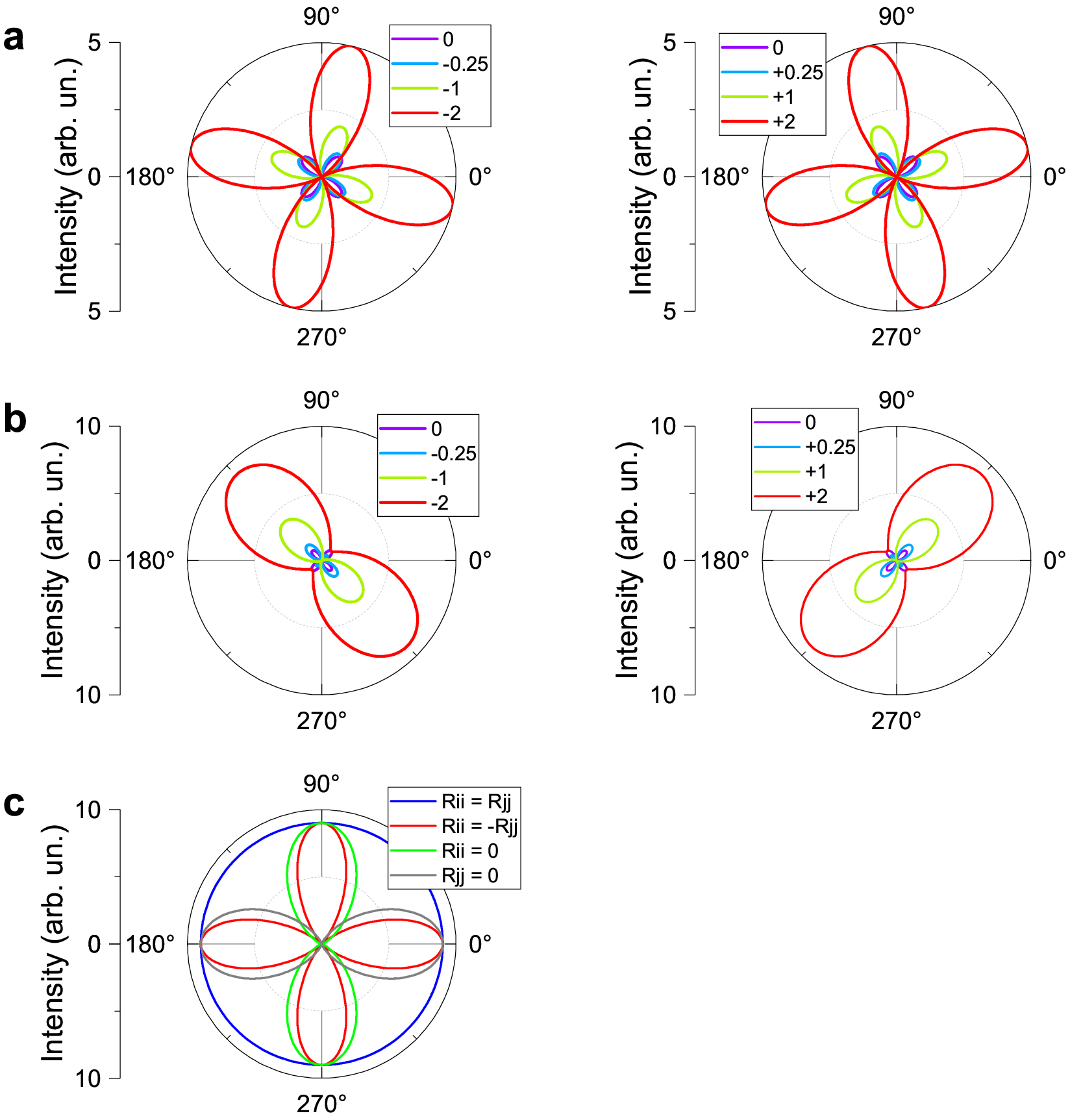}
\caption{\textbf{Suppl. Fig. 12: Effect of individual Raman tensor elements on mode symmetry.} \textbf{a}, Plots of Higgs mode intensity probed in crossed configuration based on $R_{\mathrm{axial}}$, considering fixed values for the diagonal Raman tensor elements and symmetric off-diagonal tensor elements. The labels indicate the chosen values of the off-diagonal tensor elements (in arbitrary units). \textbf{b}, The same as in a, but considering antisymmetric off-diagonal tensor elements. \textbf{c}, Higgs mode intensity probed in parallel configuration. For the diagonal tensor elements, different ratios are chosen as indicated in the label.}
\end{figure*}

To simulate the effect of varying individual Raman tensor elements, we plot several scenarios in Supplementary Fig. 12a-c. These plots are based on a general form of the Raman tensor for the Higgs mode:

\begin{center}
\mbox{$R_{\mathrm{axial}}$=$\begin{pmatrix} R_{ii} & ... & R_{ij}\\ ... & ... & ...\\ R_{ji} & ... & R_{jj}\\
\end{pmatrix}$}
\end{center}

In Supplementary Fig. 12a, we consider the Raman scattering intensity with symmetric off-diagonal tensor elements, i.e., $R_{ij} = R_{ji}$, as well as constant diagonal elements. In this case, varying the value of the off-diagonal element has no effect on the symmetry as it always remains four-fold. Instead, increasing values increase the overall intensity significantly, while also leading to a gradual counterclockwise rotation (for negative values) or clockwise rotation (for positive values). This scenario clearly does not capture our observed magnetic field response. Instead, we turn to antisymmetric off-diagonal tensor elements (i.e., $R_{ij} = -R_{ji}$), plotted in Supplementary Fig. 12b. Here, with increasing values for the off-diagonal tensor element, the symmetry gradually shifts from four-fold (in the limit $R_{ij} \to 0$) to two-fold (in the limit $R_{ij} \to \infty$). Meanwhile, a sign change flips the dominant direction by 90$^{\circ}$. Both these observations are the distinct features of our experimental field-dependent study, therefore the Raman tensor with antisymmetric off-diagonal tensor elements aptly describes our data. In Supplementary Fig. 12c we plot the Raman scattering intensity in parallel configuration for antisymmetric off-diagonal tensor elements and different ratios for the diagonal tensor element values. In cases of $R_{ii} = R_{jj}$ and $R_{ii} = -R_{jj}$, we obtain an isotropic and a four-fold response, respectively. In contrast, our experimental data clearly shows a two-fold symmetric response. This is achieved for $R_{ii} \neq R_{jj}$. In particular, in the limits $R_{ii} \ll R_{jj}$ and $R_{ii} \gg R_{jj}$ a fully two-fold symmetric response emerges.

\begin{figure*}
\label{figure17}
\centering
\includegraphics[width=16cm]{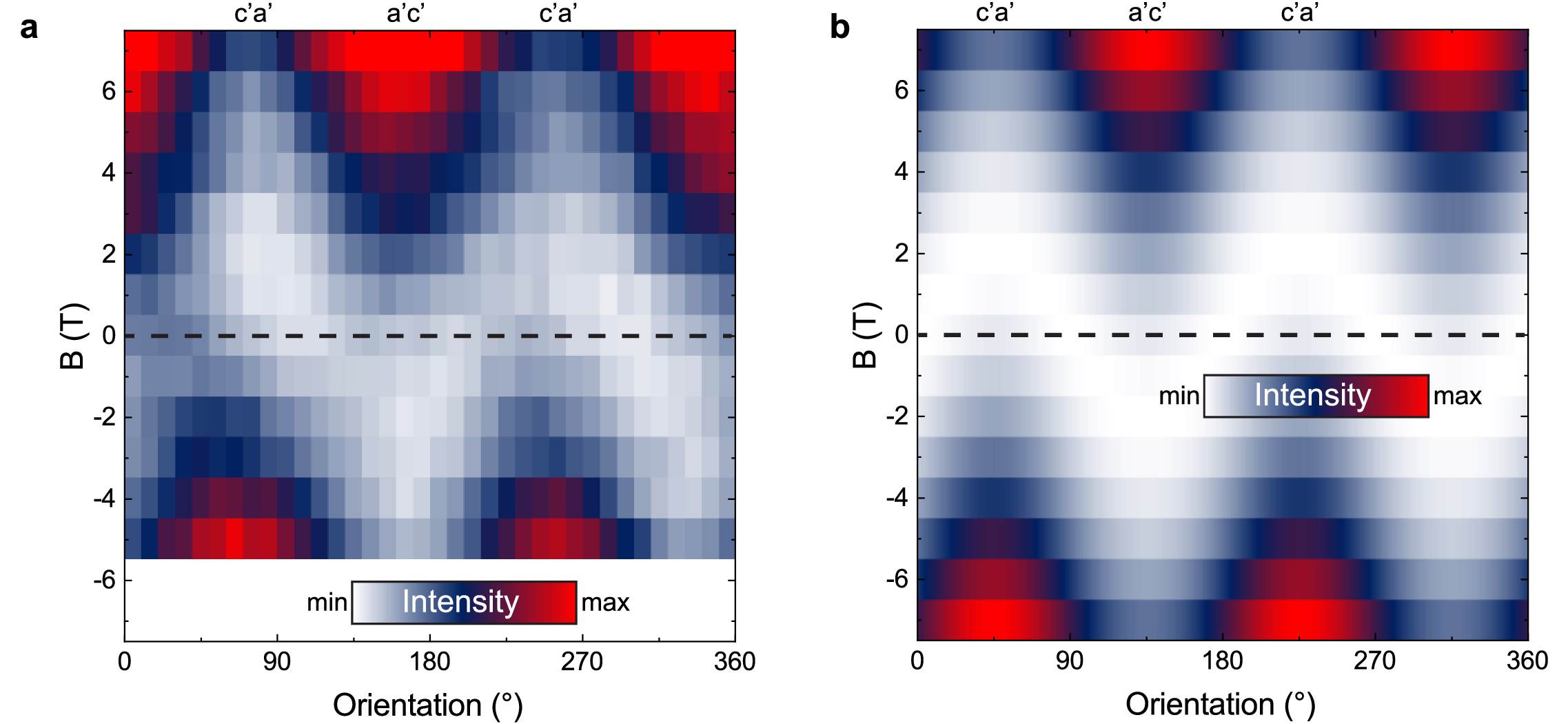}
\caption{\textbf{Suppl. Fig. 13: Field-dependent Higgs-mode intensity extracted from experiment vs. theory.} \textbf{a}, Intensity of the Higgs-type CDW amplitude mode at 9.1 meV extracted from crossed-polarization Raman scattering experiments as a function of light polarization ($x$-scale) and applied out-of-plane magnetic field ($y$-scale). \textbf{b}, Simulated Raman scattering intensity of the Higgs-mode based on $R_{\mathrm{axial}}$, considering a linear relationship between the size of the off-diagonal Raman tensor element and the applied magnetic field B.}
\end{figure*}

Based on the Raman tensor $R_{\mathrm{axial}}$ with finite, field-independent diagonal tensor elements, and anti-symmetric off-diagonal elements which depend linearly on the applied field $B$ ($R_{ij} = -R_{ji} \sim B$), we find a remarkable agreement between experiment (Supplementary Fig. 13a) and theory (Supplementary Fig. 13b), thereby fully justifying our choice for $R_{\mathrm{axial}}$.

\textbf{Supplementary Note 9 $|$ Effect of antiferromagnetic order on Raman-active modes in GdTe$_3$}

\begin{figure*}
\label{figure18}
\centering
\includegraphics[width=16cm]{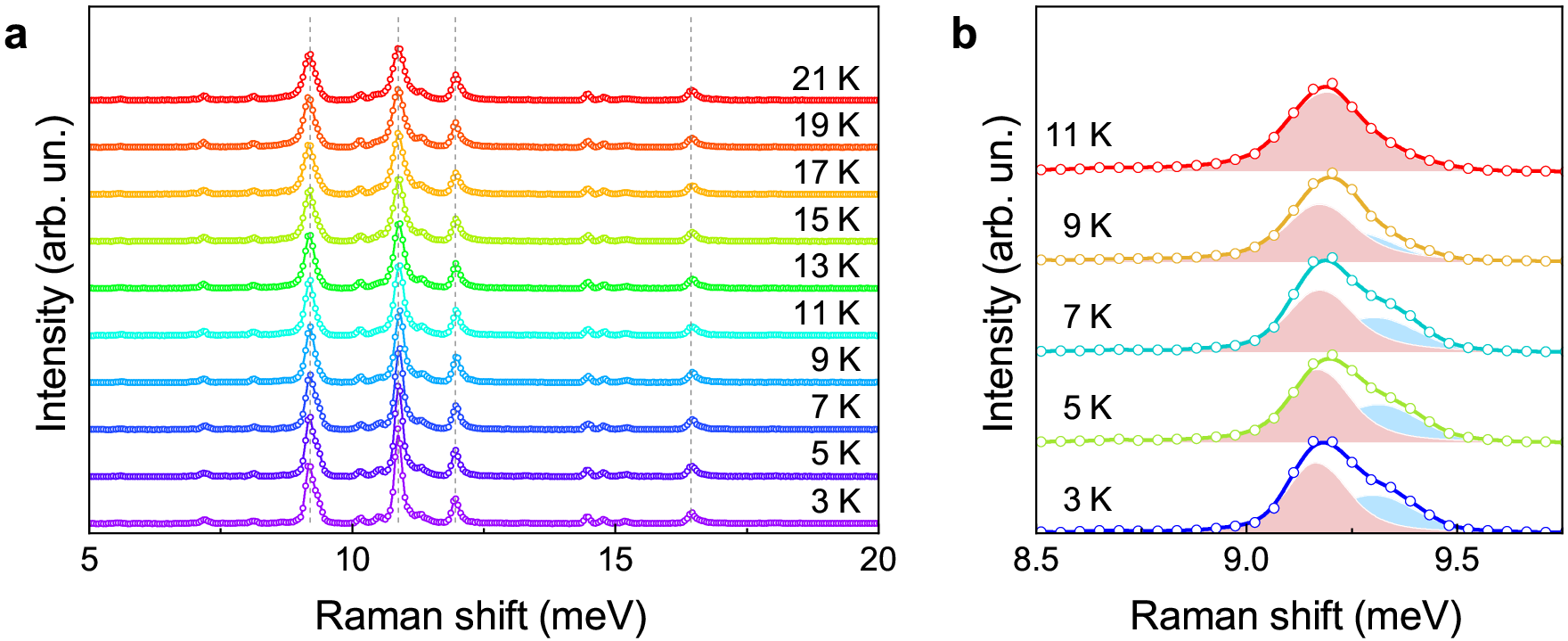}
\caption{\textbf{Suppl. Fig. 14: Raman spectra of GdTe$_3$ through $T_{\mathrm{N}}$.} \textbf{a}, Raman-active modes within an energy range of 5 meV -- 20 meV taken between 3 K and 21 K in parallel polarization. \textbf{b}, Zoom-in around the CDW amplitude mode (red shading) for temperatures up to $T_{\mathrm{N}}$. Blue shading denotes an emerging shoulder at low temperatures.}
\end{figure*}

To further confirm that the emergent antiferromagnetic order in GdTe$_3$ below $T_{\mathrm{N}} = 11.5$ K does not significantly alter the charge-density wave order, we present temperature dependent Raman spectra in Supplementary Fig. 14a. These spectra have been recorded in parallel polarization. Note that no new phonon modes appear within the AFM phase, nor do we observe any shift in frequency. In Supplementary Fig. 14b we zoom into energy range of the 9.1 meV amplitude mode, shaded in red. Here, we do observe a minor temperature dependence, namely the emergence of a weak high-energy shoulder below about 11 K, shaded in blue. We take this as a subtle fingerprint for a finite coupling between the charge-density wave and the long-range antiferromagnetic order. Importantly, however, the energy of the amplitude mode is not affected by magnetic order. Based on this observed temperature dependence, we can also further conclude that the incident laser did not result in any significant local heating of the sample during our experiments.

\textbf{Supplementary Note 10 $|$ Non-magnetic LaTe$_3$}

\begin{figure*}
\label{figure19}
\centering
\includegraphics[width=16cm]{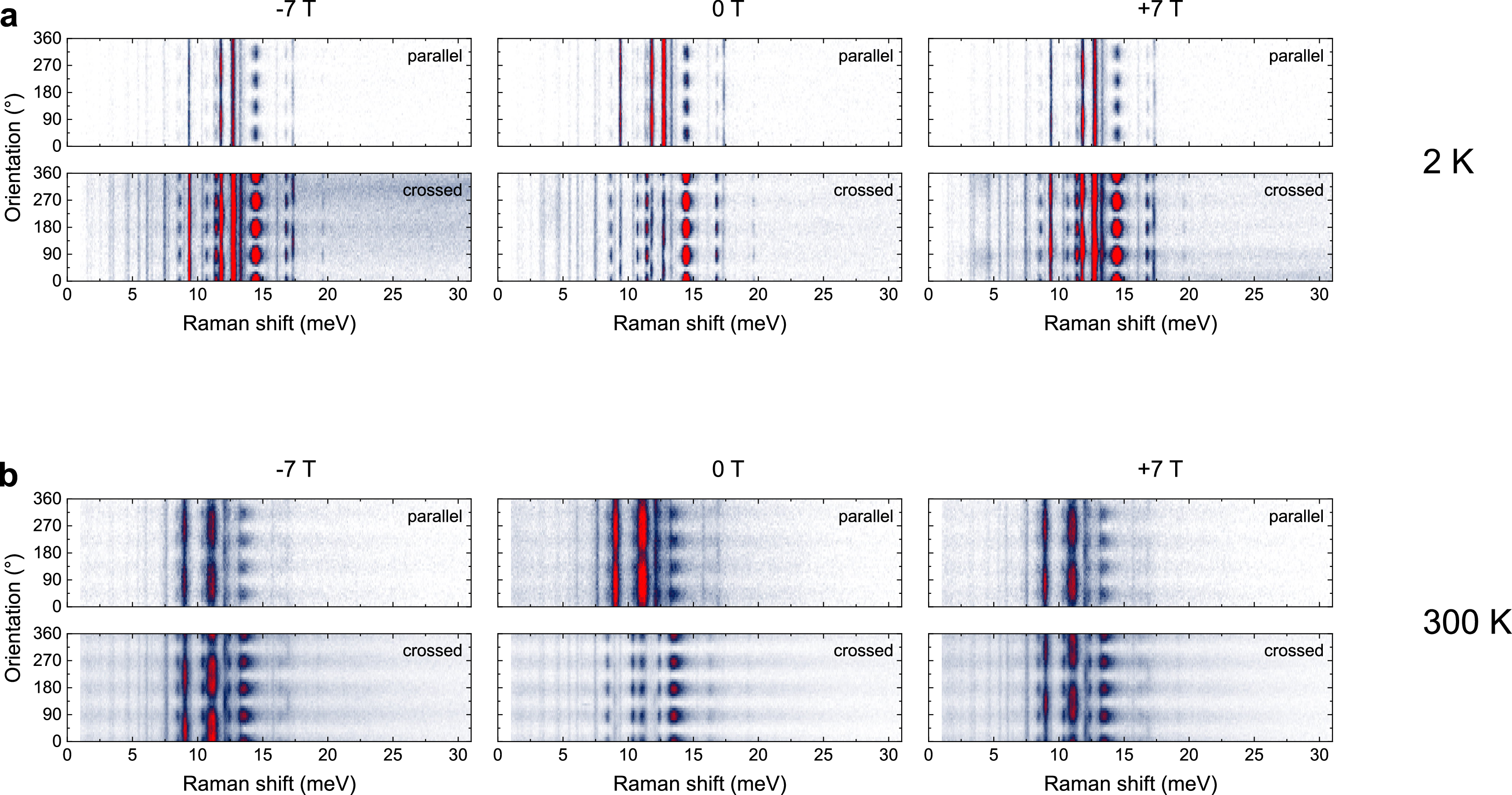}
\caption{\textbf{Suppl. Fig. 15: Field-tuning of the Higgs mode in non-magnetic LaTe$_3$.} \textbf{a}, Polarization-resolved Raman spectra of LaTe$_3$ measured at $T = 2$ K in crossed and in parallel polarization obtained at out-of-plane magnetic fields of -7 T, 0 T, and +7 T. \textbf{b}, Same measurements as in (a), obtained at $T = 300$ K.}
\end{figure*}

As a consistency check, we performed polarization-resolved Raman scattering experiments with applied out-of-plane magnetic fields on the related LaTe$_3$, which forms uniaxial charge-density-waves akin to GdTe$_3$, but lacks antiferromagnetic order at low temperatures. In Supplementary Fig. 15a (15b) results obtained at 2 K (300 K) are shown. Two CDW-Higgs modes around 11.5 and 12.5 meV are clearly seen, together with a phonon just below 15 meV. The phonon shows perfect fourfold symmetry independent of applied magnetic fields and temperature. Meanwhile, the Higgs modes assume a twofold symmetry in parallel polarization, which is also independent of magnetic field and temperature. In contrast, in crossed polarization the Higgs modes appear as intense two-fold symmetric excitation at large magnetic fields, with a 90$^{\circ}$ phase shift between positive and negative fields, while at 0 T their intensity is strongly diminished and the symmetry more ambiguous. These results are fully consistent with our observations in GdTe$_3$ and emphasize the effect out-of-plane magnetic fields have on the off-diagonal Raman tensor elements, independent of the sample's spin degrees of freedom.

\textbf{Supplementary Note 11 $|$ Absence of chiral phonons in GdTe$_3$}

\begin{figure*}
\label{figure20}
\centering
\includegraphics[width=8cm]{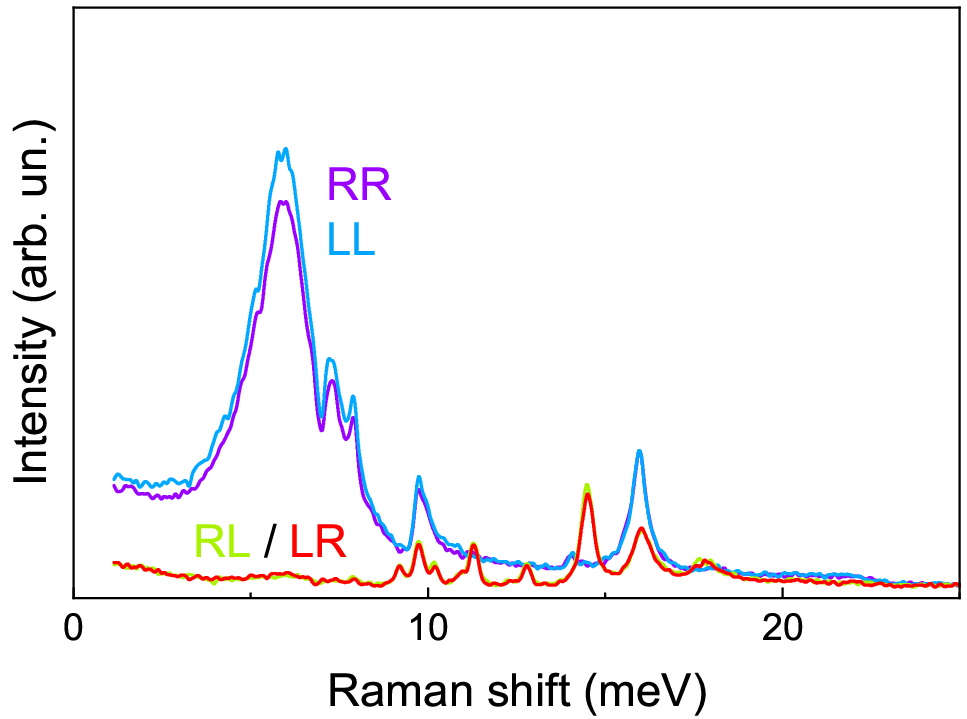}
\caption{\textbf{Suppl. Fig. 16: Raman spectra of GdTe$_3$ taken with circularly polarized light in parallel ($RR$, $LL$) and in crossed ($RL, LR$) configuration at $T = 300$ K and $B = 0$ T.}}
\end{figure*}

Supplementary Fig. 16 shows Raman spectra obtained using circularly polarized light, with $R = x+\mathrm{i}y$ and $L = x-\mathrm{i}y$. In case of spontaneous ferro-rotational order, chiral phonons may be anticipated as an experimental fingerprint, which would result in a subtle imbalance between Raman scattering intensity probed in $RL$ and $LR$ configuration, and possibly a splitting of phonon energies as a consequence of lifted degeneracies for certain modes~\cite{ishito-23}. Within our resolution, we do not observe any splitting or intensity imbalance. Although this may not strictly rule out the existence of chiral phonons and, consequently, ferro-rotational order in GdTe$_3$, we do not have any experimental support for such a scenario.

\textbf{Supplementary Note 12 $|$ Field-poling of domains vs field-enhanced axiality}

\begin{figure*}
\label{figure21}
\centering
\includegraphics[width=10cm]{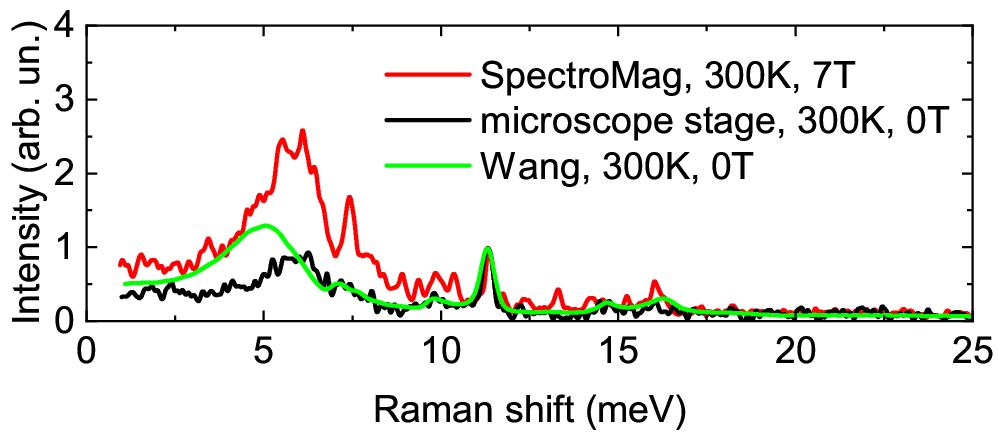}
\caption{\textbf{Suppl. Fig. 17: Comparison of Raman spectra measured at room temperature in (a'c') polarization (i.e., in-plane crossed, where the amplitudon around 5 meV is maximized).} Red line: Data taken within the magneto-optical cryostat with a large laser spot diameter ($\sim$ 100 $\mu$m) and an applied field of 7 T. Black line: Data taken using a microscope stage with a tight laser spot diameter ($\sim$ 2 $\mu$m) at $B = 0$. Green line: Reference data~\cite{wang-22} with a tight laser spot diameter ($\sim$ 2 $\mu$m) at $B = 0$.}
\end{figure*}

In this Supplementary Note we compare the Higgs mode intensity measured without applied field and across a tight spot, i.e., on a putative single domain (Supplementary Fig. 17, green curve and black curve) to the intensity measured across a large area with 7 T applied (Supplementary Fig. 17, red curve). We find that magnetic fields lead to a substantial surplus in Higgs-mode intensity. This is also in line with the linear and continuously increasing Higgs mode intensity upon applying large magnetic fields. Therefore, we conclude that in addition to poling the domains, which takes place dominantly at low fields, magnetic fields also affect the OAM itself. To facilitate this direct comparison between measurements carried out in different conditions and in different groups we normalized all spectra to the intensity of the 12 meV phonon. The difference in Higgs mode energy between our spectra (red and black curves) and the reference curve (green) might stem from slightly different sample temperatures, since close to $T_{CDW}$ the Higgs mode starts to soften dramatically with increasing temperature.

\textbf{Supplementary Note 13 $|$ Experimental Setup}

\begin{figure*}
\label{figure22}
\centering
\includegraphics[width=16cm]{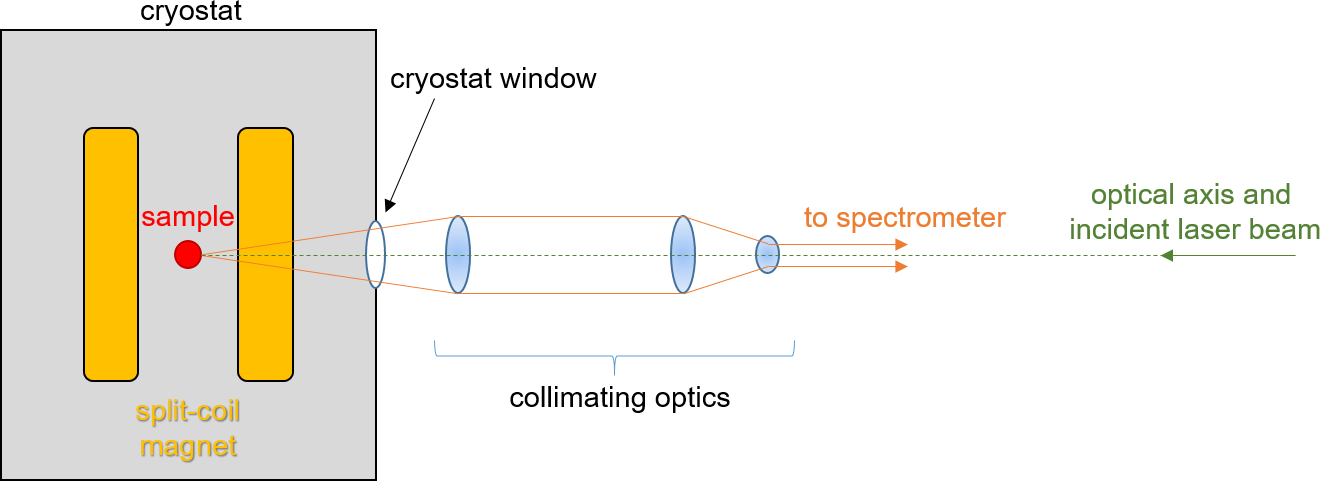}
\caption{\textbf{Suppl. Fig. 18: Optical beampath to and from the sample inside the magneto-optical cryostat.}}
\end{figure*}

To further address the potential issue of Faraday rotation, we show the detailed experimental setup in Supplementary Fig. 18. Here, all focusing and collecting optics are located outside the cryostat, with the first lens placed 180 mm away from the sample and from the center of the magnetic field. Note that outside the cryostat the maximum magnetic stray field is below 1 T (measured closest to the cryostat at a maximum applied center field of 7 T). Therefore we believe that any existing field-induced Faraday rotation would be insignificantly small. Furthermore, we did not observe any continuous phase shift of the periodic intensity profiles of excitations as a function of light polarization with increasing magnetic field (see Supplementary Fig. 5). As the magnetic field is applied and increased, the maxima and minima remain at their fixed angles, underlining the absence of any major Faraday rotation. We also recall that while some of the CDW-coupled and zone-folded phonons do exhibit anomalous intensity modifications as a function of magnetic field, the effect is less pronounced or distinct from what we observe for the CDW amplitudon, i.e., different excitations react differently to applied fields. All of these observations suggest that our reported CDW anomaly stems from intrinsic physics rather than extrinsic artifacts such as Faraday rotation.

\end{document}